\documentclass[11pt,a4paper]{article}
\usepackage{amssymb,amsmath,amsfonts,amsthm,amstext,amscd,array}
\usepackage{mathrsfs}
\usepackage{bbm}
\usepackage{bm}
\usepackage[arrow,matrix,curve]{xy}
\usepackage{bbding}
\usepackage{wasysym}

\def\tri{\triangleright}
\def\rbtri{\blacktriangleright}
\def\lbtri{\blacktriangleleft}

\newcommand{\eps}{\varepsilon}

\def\ii{{\,{\rm i}\,}}
\def\dd{{\rm d}}

\def\Id{{\rm id}}

\def\mfg{{\mathfrak g}}
\def\mfh{{\mathfrak h}}

\def\mcH{{\mathcal H}}
\def\mcF{{\mathcal F}}
\def\mcR{{\mathcal R}}

\newcommand{\CCC}{\mathscr{C}}

\newcommand{\CCF}{\mathscr{F}}
\newcommand{\CCM}{\mathscr{M}}

\newcommand{\eq}{\begin{equation}}
\newcommand{\eqend}{\end{equation}}
\newcommand{\eqa}{\begin{eqnarray}}
\newcommand{\nonueqa}{\begin{eqnarray*}}
\newcommand{\eqaend}{\end{eqnarray}}
\newcommand{\nonueqaend}{\end{eqnarray*}}

\newcommand{\bma}[1]{\begin{array}{#1}}
\newcommand{\ema}{\end{array}}
\newcommand{\bc}{\begin{center}}
\newcommand{\ec}{\end{center}}

\renewcommand{\thefootnote}{\fnsymbol{footnote}}
\newcommand{\newsection}{\setcounter{equation}{0}\section}

\newcommand{\complex}{{\mathbb C}} 
\newcommand{\zed}{{\mathbb Z}} 
\newcommand{\real}{{\mathbb R}} 

\newcommand{\torus}{{\mathbb T}} 

\def\hil{{\mathcal H}}

\def\Mcal{{\mathcal M}}
\def\Tcal{{\mathcal T}}
\def\ot{{\, \otimes\, }}

\newif\ifold             \oldtrue

\def\e{{\,\rm e}\,}

\def\be{\begin{equation}}
\def\ee{\end{equation}}
\def\bea{\begin{eqnarray}}
\def\eea{\end{eqnarray}}
\def\bd{\begin{displaymath}}
\def\ed{\end{displaymath}}

\def\s{\sigma}

\newcommand{\beq}{\begin{eqnarray}}
\newcommand{\eeq}{\end{eqnarray}}

\makeatletter
\newdimen\normalarrayskip              
\newdimen\minarrayskip                 
\normalarrayskip\baselineskip
\minarrayskip\jot
\newif\ifold             \oldtrue            
\def\arraymode{\ifold\relax\else\displaystyle\fi} 
\def\@arrayskip{\ifold\baselineskip\z@\lineskip\z@
     \else
     \baselineskip\minarrayskip\lineskip2\minarrayskip\fi}
\def\@arrayclassz{\ifcase \@lastchclass \@acolampacol \or
\@ampacol \or \or \or \@addamp \or
   \@acolampacol \or \@firstampfalse \@acol \fi
\edef\@preamble{\@preamble
  \ifcase \@chnum
     \hfil$\relax\arraymode\@sharp$\hfil
     \or $\relax\arraymode\@sharp$\hfil
     \or \hfil$\relax\arraymode\@sharp$\fi}}
\def\@array[#1]#2{\setbox\@arstrutbox=\hbox{\vrule
     height\arraystretch \ht\strutbox
     depth\arraystretch \dp\strutbox
     width\z@}\@mkpream{#2}\edef\@preamble{\halign \noexpand\@halignto
\bgroup \tabskip\z@ \@arstrut \@preamble \tabskip\z@ \cr}%
\let\@startpbox\@@startpbox \let\@endpbox\@@endpbox
  \if #1t\vtop \else \if#1b\vbox \else \vcenter \fi\fi
  \bgroup \let\par\relax
  \let\@sharp##\let\protect\relax
  \@arrayskip\@preamble}
\makeatother

\newcommand{\wg}{\wedge}

\newcommand{\p}{\partial}
\newcommand{\mc}{\mathcal}

\newcommand{\trm}{\textrm}

\newcommand{\scr}{\scriptsize}

\def\tmfg{{\mfg}_e}

\def\tF{{\bar F}}
\def\tP{{\tilde P}}
\def\tp{{\bar p}}
\def\tM{{\bar M}}

\def\tmcF{{\bar \mcF}}
\def\As{{\APLstar}}
\def\tAs{{\,\bar \As \,}}
\def\ts{{\, \bar \star \,}}
\def\tmcR{{\bar \mcR}}
\def\moyal{{\, \mbox{{\tiny \FourStar}}\,}}

\theoremstyle{definition}

\newcommand{\intps}{\int_{{\cal M}} \, \dd^{2d}x \ }
\def\cirp{\mathop{\bar\circ}}

\usepackage{hyperref}
\hypersetup{colorlinks,%
citecolor=black,%
filecolor=black,%
linkcolor=black,%
urlcolor=black,%
unicode,pdftitle={Nonassociative twist
  deformations},pdfauthor={Dionysios Mylonas, Peter Schupp and Richard Szabo}}

\begin{document}
\begin{titlepage}
\begin{flushright}

\baselineskip=12pt

EMPG--13--20

\end{flushright}

\begin{center}

\vspace{2cm}

\baselineskip=24pt

{\Large\bf Non-Geometric Fluxes, Quasi-Hopf Twist Deformations \\ and
  Nonassociative Quantum Mechanics}

\baselineskip=14pt

\vspace{1cm}

{\bf Dionysios Mylonas}${}^{1}$ \footnote{Email: \ {\tt mylonas.dionysios@gmail.com}}, {\bf
  Peter Schupp}${}^{2}$ \footnote{Email: \ {\tt
    p.schupp@jacobs-university.de}}  \ and \ {\bf Richard
  J. Szabo}${}^{1}$ \footnote{Email: \ {\tt R.J.Szabo@hw.ac.uk}}
\\[5mm]
\noindent  ${}^1$ {\it Department of Mathematics, Heriot-Watt University\\ Colin Maclaurin Building,
  Riccarton, Edinburgh EH14 4AS, U.K.}\\ and {\it Maxwell Institute for
Mathematical Sciences, Edinburgh, U.K.} \\ and {\it The Tait Institute, Edinburgh, U.K.}
\\[3mm]
\noindent ${}^2$ {\it Jacobs University
  Bremen\\ 28759 Bremen, Germany}
\\[30mm]

\end{center}

\begin{abstract}

\baselineskip=12pt

We analyse the symmetries underlying 
nonassociative deformations of geometry in
non-geometric $R$-flux compactifications which arise via T-duality
from closed strings with constant geometric fluxes. Starting from the non-abelian Lie
algebra of translations and Bopp shifts in phase space, together with
a suitable cochain twist, we construct the quasi-Hopf algebra of
symmetries that
deforms the algebra of functions and the exterior differential calculus in the
phase space description of nonassociative
$R$-space. In this setting nonassociativity 
is characterised by the associator 3-cocycle which controls 
non-coassociativity of the quasi-Hopf algebra. We use abelian 2-cocycle twists to
construct maps between the dynamical
nonassociative star product and a family of associative star products parametrized by constant momentum
surfaces in phase space. 
We define a suitable integration on these nonassociative
spaces and find that the usual cyclicity of associative noncommutative
deformations is replaced by weaker notions of $2$-cyclicity and
$3$-cyclicity. Using this star product quantization on phase space
together with 3-cyclicity, we
formulate a consistent version of nonassociative quantum mechanics, in
which we calculate the expectation values of area and volume
operators, and find coarse-graining of the string background due to the $R$-flux.

\end{abstract}

\end{titlepage}
\setcounter{page}{2}

\newpage

{\baselineskip=12pt
\tableofcontents
}

\bigskip

\renewcommand{\thefootnote}{\arabic{footnote}}
\setcounter{footnote}{0}


\newsection{Introduction and summary \label{intro}}

Flux compactifications of superstring theory (see e.g.~\cite{Grana,Douglas:2006es,BKLSrev} for reviews)
enjoy T-duality symmetry which in some instances map them to non-geometric
spaces~\cite{Shelton:2005cf}, and in this way non-geometric flux backgrounds can also arise as
consistent closed string vacua. In this paper we consider the emergence of
non-geometry under the action of T-duality on torus fibrations with fluxes. In particular, the geometry
probed by closed strings propagating in a flat three-torus $\torus^3$
endowed with a constant Neveu-Schwarz $H$-flux can be regarded as
the identity fibration $\torus^3\to\torus^3$ with zero-dimensional
fibres. It is mapped under
T-duality along a cycle to a circle bundle over $\torus^2$ of degree
equal to the cohomology class of the $H$-flux in $H^3(\torus^3;\zed)=\zed$; this is
the twisted three-torus with zero flux but non-vanishing metric
torsion. A further T-duality along a cycle of the base yields a
$\torus^2$-fibration over $S^1$, which is a non-geometric space with
$Q$-flux (the flux dual to the $H$-flux); this space is an example of
a T-fold~\cite{Hull:2004in}, which is locally Riemannian but globally
the transition functions include T-duality transformations in addition
to the standard diffeomorphisms. As a result of non-trivial $SL(2,\zed)$
monodromies of the $\torus^2$ fibres induced by wrapping closed strings around the
$S^1$ base, the fibre directions acquire a noncommutative deformation
induced by both the $Q$-flux and the winding number along the base
direction~\cite{Lust:2010iy,Condeescu:2012sp,Andriot:2012vb,Hassler:2013wsa}.\footnote{Non-geometric $Q$-flux backgrounds dual to three-spheres were also
recently constructed in~\cite{Plauschinn:2013wta}.}
 After a final T-duality of the base $S^1$ one obtains a
$\torus^3$-fibration over a point. Then the $Q$-flux is mapped to its
T-dual $R$-flux while the winding number is mapped to the string
momentum, thus yielding the quantum phase space algebra~\cite{Lust:2010iy}
\be 
[x^i,x^j]=\ii \hbar\, R^{ijk} \,p_k \qquad ,\qquad [x^i,p_j]= \ii \hbar \, \delta^i{}_j \qquad\mbox{,} \qquad [p_i,p_j]=0 \ , \label{luestalg}
\ee
where $x^i,\, p_i$ are the zero modes of the string positions and
conjugate momenta, respectively;\footnote{Throughout this paper we use
units in which Planck's constant $\hbar$ is dimensionless. We also use
implicit summation over repeated upper and lower indices throughout.} in a general 
$d$-dimensional non-geometric parabolic $R$-flux
background, which is the main focus of this paper, the transition
functions cannot even be defined locally and a short calculation
reveals that it also acquires a \emph{nonassociative} deformation since the algebra \eqref{luestalg} has a non-trivial Jacobiator given by
\be
[[x^i,x^j,x^k]] := \big[[x^i,x^j],x^k\big]+\mbox{cyclic permutations} = 3 \, \hbar^2 \, R^{ijk} \ . \label{jacobiator}
\ee
In the following we also consider suitable decompactification limits
of the parabolic $R$-flux model to spaces with trivial topology.

Non-geometric backgrounds can also be studied in the context of
worldsheet conformal field theory. In this setting geometric flux
backgrounds correspond to conformal field theories on freely acting
orbifolds with left-right
symmetric twisted sectors with respect to the action of the
$SL(2,\zed)$ monodromy
on $\torus^2$, while
non-geometric backgrounds are regarded as left-right asymmetric orbifold
theories whose asymmetric twisting is related to the presence of non-geometric
fluxes~\cite{Condeescu:2012sp, Condeescu:2013yma}. Such conformal
field theories exhibit the nonassociative structure of the underlying
target space as a discontinuity of the three-point
functions~\cite{Blumenhagen:2010hj}. In fact, one can read off a
deformed product of three functions up to linear order in the
background flux by calculating off-shell correlation functions of
tachyon vertex operators~\cite{Blumenhagen:2011ph}. 

From a target
space perspective, a suitable framework for describing non-geometry
in string theory is provided by both double field
theory~\cite{Hull:2009mi}  (see e.g.~\cite{Aldazabal:2013sca,Berman:2013eva,Hohm:2013bwa} for
reviews) and generalized geometry (see
e.g.~\cite{Grana:2008yw}). Double field theory implements $O(d,d)$
transformations as a symmetry of the string theory
effective action via a doubling of the dimension of space through the
introduction of
dual coordinates on equal footing with the original space coordinates; hence it encompasses all T-duality frames in an invariant way,
including those exhibiting closed string noncommutativity and nonassociativity. On the other hand, generalized geometry
amounts to the extension of differential geometry to algebroids which
accommodates string symmetries, through an $O(d,d)$-invariant bilinear
form on sections of the algebroid. These
approaches can be used in a complimentary way to construct string
actions in non-geometric frames in which diffeomorphism and gauge
symmetries are expressed via generalised
geometry~\cite{Andriot:2012an}. In fact it is possible to
formulate a
bi-invariant action, i.e. invariant under both diffeomorphisms and
$\beta$-transformations of the algebroid, for closed strings in non-geometric flux
backgrounds~\cite{Blumenhagen:2012nt}; in particular, each $O(d,d)$
transformation is associated with a Lie
algebroid~\cite{Blumenhagen:2013aia}. In addition to providing the
appropriate geometric formalism for describing non-geometric
backgrounds,
these
methods go a step further in providing a suitable context for a
desired deformation quantization of these spaces which is not
present in the usual description of closed string theory; the
structure of nonassociative deformations of geometry in double field
theory is analysed in~\cite{Blumenhagen:2013hva}. {Some of these
  non-geometries can also be equivalently
  described 
  entirely within the geometric framework of the ten-dimensional supergravity
  theory proposed by~\cite{Andriot:2013xca}, without recourse to any
  worldsheet formalism or the target space
  formalisms based on double field theory and generalized geometry.}

An alternative geometric interpretation of the non-geometric $R$-flux background
along these lines was given in~\cite{Mylonas:2012pg} in terms of
certain 
membrane sigma-models, proposed as topological sectors of closed
string dynamics in flux compactifications. Courant algebroids provide the
appropriate target spaces for these sigma-models in order to
accomodate the 3-tensor fluxes. In particular,  for constant fluxes the topological
sigma-model on the standard
Courant algebroid $TM\oplus T^*M$ (in a suitable frame for the
$\beta$-transformation symmetry) over a manifold $M$ of dimension $d$
reduces
on the (closed string) boundary of the membrane to
sigma-models whose target spaces are Lie algebroids over twisted Poisson
manifolds. We showed in~\cite{Mylonas:2012pg} that the geometric
$H$-flux background corresponds in this way to a twisted Poisson
sigma-model with target space $M$, while the non-geometric $R$-flux
background corresponds to a twisted Poisson sigma-model whose target
space is the {cotangent bundle} ${\cal M}=T^*M$, thus reproducing the
twisted Poisson structure on phase space which is quantized by (\ref{luestalg}).

Based on an open/closed string duality in $R$-space arising from the
asymmetric twisted sectors of the orbifold conformal field
theory~\cite{Mylonas:2012pg}, this interpretation enables an explicit construction of a
nonassociative star product on the algebra of functions
$C^\infty({\cal M})$ via two techniques: Kontsevich's deformation
quantization of a twisted Poisson manifold~\cite{Konts:1997} and strict deformation quantization of the dual of a Lie
2-algebra via convolution in an integrating Lie 2-group. The first
method is related to the quantization of Nambu-Poisson structures on
the original compactification manifold $M$,
while the second method leads to a categorification of Weyl's
quantization map and clarifies the relation with the twisted
convolution products on nonassociative torus
bundles~\cite{Bouwknegt:2004ap}. Equivalence of the two methods was
demonstrated by a categorified version of Kathotia's theorem which asserts the
equivalence between Baker-Campbell-Hausdorff quantization and
Kontsevich's deformation quantization in the case of nilpotent Lie
algebras~\cite{Kath:1998}. In particular, the non-trivial associator calculated
in~\cite{Mylonas:2012pg} is determined by a classifying 3-cocycle in the
Chevalley-Eilenberg cohomology with values in the trivial
representation of the $d$-dimensional Heisenberg group
which integrates the phase space algebra in the $Q$-flux duality
frame. This 3-cocycle
provides the twisting of the pertinent horizontal product of the Lie
2-group, which was calculated explicitly for the case of toroidal
backgrounds using the Baker-Campbell-Hausdorff formula. At the
algebraic level, this 3-cocycle is exactly the one that appeared
recently in~\cite{Bakas:2013}, where $R$-space nonassociativity was similarly
characterised in
terms of 3-cocycles of the abelian group of translations in phase
space with and without
central extension using the deformation theory of Lie algebras (see also~\cite{Castellani:2013mka}).

In this paper we describe a third way of quantizing non-geometric
$R$-flux backgrounds using twist deformation techniques. The
terminology ``twist'' refers to a deformation of a {Hopf algebra} $H$
which is constructed from the universal enveloping algebra of a Lie
algebra of symmetries acting on the phase space description of
$R$-space. Such deformations are typically provided by a 2-cocycle
$F\in H \ot H$
called a \emph{twist} (see e.g.~\cite{Majid:book}); gauge and gravity
theories on a noncommutative space as deformations of their classical
counterparts using twisting techniques can be
found e.g. in~\cite{Aschieri:2005yw,Aschieri:2006ye} (see also~\cite{Aschieri:2012ii}). The advantage of
twist deformation quantization is that it accommodates
{nonassociativity} in a natural and concrete way which
overcomes the difficulties encountered in quantizing nonassociative
algebras using (higher) Lie algebraic methods, such as
Baker-Campbell-Hausdorff quantization: One simply
requires that the usual coassociativity of the Hopf algebra $H$ holds only up to
a 3-cocycle $\phi\in H\otimes H\ot H$ called the \emph{associator};
this yields a quasi-Hopf algebra~\cite{Dri89}. If $\phi$ is trivial,
i.e. it is the coboundary of a 2-cochain $F\in H\otimes H$, then the twisting is
provided by $F$. 
Once a twist is known, it is just a matter of applying the cochain
twist machinery to
deform all geometric structures which are covariant under the symmetries of a manifold; such cochain twists
were employed in~\cite{Beggs:2010} to describe nonassociative
differential calculus and in~\cite{Majid:2005} to formulate gauge
theory on nonassociative algebras (see also~\cite{Beggs:2010b}). 

The use of trivial 3-cocycles as sources of
  nonassociativity first appeared in the physics literature in the
  description of magnetic translations of charged particles in the background of a magnetic
  monopole, where it was shown that demanding associativity yields
  the Dirac quantization condition~\cite{Jackiw:1984rd}. In this case
  one finds an associative representation of the global translations, even
  though the Jacobi identity for the infinitesimal generators
  continues to fail. This point of view is taken in the description of
non-geometric toroidal flux backgrounds within the framework of Matrix
theory compactifications in~\cite{Chatzistavrakidis:2012qj}. Thus although one finds a
non-trivial Jacobiator (\ref{jacobiator}) on $R$-space, demanding
associativity of global quantities may teach us something about
the structure of non-geometric fluxes, such as flux
quantization; indeed, the on-shell worldsheet tachyon scattering
amplitudes computed
in~\cite{Blumenhagen:2011ph} exhibit no violations of associativity once
momentum conservation is taken into account, in accord with the
standard crossing symmetry of correlation functions in two-dimensional
conformal field theory. This point of view of nonassociative $R$-space
is addressed in the context of double field theory
in~\cite{Blumenhagen:2013hva}, while the parallels between
nonassociative parabolic $R$-flux string models and the dynamics of charged particles
in uniform magnetic charge distributions is elucidated in~\cite{Bakas:2013}.

Twist deformations can also be formulated in a categorical framework,
based on the observation that every braided monoidal category whose
objects are vector spaces is equivalent to the representation category of
some (quasi-)Hopf algebra $H$ (see
e.g.~\cite[Chap.~9]{Majid:book} for details). The associator in this
case is a functorial isomorphism which satisfies the pentagonal
coherence relations in the braided monoidal category of left
$H$-modules. This categorical formulation connects the strict
deformation quantization techniques developed in~\cite{Mylonas:2012pg}
for non-geometric $R$-flux backgrounds with the twist deformation
quantization techniques which are developed in the present paper; in
particular, it \emph{explicitly} yields the generalisation of
Kathotia's theorem to Lie 2-algebras. The
formalism of twist deformation quantization is reviewed in
Section~\ref{td}. 

In this paper we demonstrate that twist deformation quantization
allows for the introduction of a 3-form $R$-flux in phase space in
a natural way. In Section~\ref{deformation} we do this in three steps
which illustrate that the nonassociativity of $R$-space is not completely
arbitrary, in that it results from the fact that the star product is a
function of momentum, i.e. it is dynamical. We shall find a whole
family of associative star products for constant momentum slices, all
interrelated among themselves and to the nonassociative star product
by twists; these relations were described by Seiberg-Witten maps in~\cite{Mylonas:2012pg}. First we construct the Hopf
algebra $K$ related to the abelian Lie algebra of translations in
$2d$-dimensional phase space ${\cal M}$, and deform it using an
abelian 2-cocycle twist $F\in K\ot K$; the action of the twisted Hopf
algebra $K_F$ on the algebra of functions $C^\infty({\cal M})$ yields
the canonical Moyal-Weyl star product on phase space. We then endow ${\cal M}$
with a trivector $R$ which is T-dual to the 3-form of a uniform background
$H$-flux. To bring $R$ into
the twist quantization scheme, we introduce a unique family of twist
elements associated to the 
translation algebra which are parametrised by
constant momentum, and hence deform the
pointwise product of functions on phase space to a family of
associative noncommutative Moyal-Weyl type star products; these sorts of deformed products
were derived in~\cite{Mylonas:2012pg} for $Q$-flux backgrounds
(T-folds). Finally, we promote constant momenta to dynamical momenta
appropriate to $R$-space and twist the pertinent Hopf algebra $H$ using
a cochain twist, which is tantamount to an abelian cocycle twist of
the canonical Moyal-Weyl product; the underlying Lie algebra of symmetries of $R$-space
is non-abelian, nilpotent and
includes non-local Bopp shifts on ${\cal M}$ that mix positions with
momenta. The resulting twisted Hopf algebra is a quasi-Hopf algebra
whose action on $C^\infty({\cal M})$ quantizes the phase space
structure of constant $R$-flux backgrounds and yields the nonassociative star
product on $R$-space that was first proposed
in~\cite{Mylonas:2012pg}. 

Equiped with our cochain twist, in Section~\ref{calculus} we deform the differential calculus on
${\cal M}$, and thus formulate nonassociative deformations of the
exterior differential algebra and of the $C^\infty({\cal M})$-bimodule
structure on $R$-flux backgrounds. In
order to set up a framework in which to study field theories on
$R$-space, we define integration on the deformed algebra of forms on
${\cal M}$ to be the standard integration; the integral of multiple
exterior star products of differential forms is not (graded) cyclic but rather satisfies weaker
notions of 2-cyclicity and 3-cyclicity that we describe, which turn
out to be crucial for a
consistent formulation of quantum theory on $R$-space. 3-cyclicity is
also crucial for ensuring that nonassociative deformations of field
theory can be made consistent with the requirements of crossing
symmetry of conformal
field theory scattering amplitudes, which on-shell show no violations
of associativity.

Nonassociative quantum theory on non-geometric
$R$-flux backgrounds cannot be treated
by conventional means in terms of linear operators on separable
Hilbert spaces; instead, one should resort to our phase space star
product quantization (this point is also emphasised
by~\cite{Bakas:2013}). We elucidate this point in Section~\ref{NAQM} by considering 
quantum mechanics in the nonassociative phase space formalism; in
this approach we emphasise the roles of observables without concern
about their representations. We demonstrate that, against all odds, a
consistent formulation of nonassociative quantum mechanics is indeed
possible; an important input is the 3-cyclicity of the nonassociative
star product. This analysis demonstrates that quantum theory on the
nonassociative string background can be treated in a systematic way, and our treatment
is the first step towards realising more elaborate models, such as
field theory or gravity, on 
non-geometric flux compactifications. Through this
formalism we find that a triple of operators that do not associate does not
have common eigenstates, which is a clear sign of position space quantization
in the presence of $R$-flux. An uncertainty relation proportional to the transverse momentum for the
measurement of a pair of position coordinates is induced. We also find
non-zero expectation values for (the uncertainty of) suitably defined
area and volume operators in configuration space, leading in particular to a minimal volume
element; this formalism thus provides a concrete and rigorous
derivation for the uncertainty relations anticipated by~\cite{Lust:2010iy,Blumenhagen:2010hj}. We shall further find that operator time
evolution in the Heisenberg picture is not a derivation of the star product algebra of
operators.

Finally, some generalizations of our twist deformation methods to
non-constant $R$-flux backgrounds as well as more generic $R$-flux
string vacua are briefly considered in Section~\ref{nonconstant}. We
consider the case of position-dependent $R$-fluxes in (\ref{luestalg})
and the conditions under which the techniques of twist deformation quantization
developed in this paper carry through at least locally, leaving the
difficult problem of globalization, which is analogous to that of the star products in Kontsevich's approach~\cite{Cattaneo:2001vu}, to future work; in
particular, by restricting to functions of the position coordinates in
$M$, this technique provides a framework for quantizing generic
Nambu-Poisson 3-brackets determined by the trivector field $R$. We
also consider the extension of the phase space algebra
(\ref{luestalg}) to allow for twist deformations provided by
a class of quasi-Poisson structures that are generic
non-linear functions of the momenta, which is the case of $R$-flux
backgrounds that arise from monodromies
lying in certain non-parabolic conjugacy classes of the $SL(2,\zed)$
automorphism group of $\torus^2$. In this
case we apply Kontsevich's deformation quantization of phase
space to compute the nonassociative star product and associator up to third order in a derivative expansion in the $R$-flux,
which is used to identify the pertinent Hopf algebra of symmetries and a
(non-unique) cochain twist. The generality of this setting allows for
deformation quantization of closed string $R$-flux backgrounds
determined by the elliptic model of~\cite{Lust:2010iy,Condeescu:2012sp} and illustrates
the power of our twist deformation techniques: The passage to these
non-parabolic $R$-flux models amounts to extending the 
Lie algebra of translations and Bopp shifts in phase space to a certain
infinite-dimensional Lie algebra of diffeomorphisms.


\newsection{Twist deformation quantization \label{td}}

Twist deformation techniques provide a very precise and systematic way of quantizing any algebraic structure acted upon by a (quasi-)Hopf algebra. Such is the case for the algebra of functions on a space acted upon by a Lie group of symmetries which will be our main application in this paper. In this section we briefly review standard deformation quantization by cocycle twists as well as the more general case of cochain twists which is our main case of interest.


\subsection{Hopf algebras and cocycle twist quantization\label{ha}}

We begin by defining some of the basic algebraic structures that we will encounter in the following. We then describe deformation quantization via a Drinfel'd cocycle twist.

A \emph{bialgebra} $H$ over $\complex$ is an associative unital algebra with a counital coalgebra structure that satisfies the properties
\eqa
&(\Id_H\ot\Delta) \circ\Delta = (\Delta\ot\Id_H)\circ\Delta \ , & \label{coassoc}\\[4pt]
&(\Id_H\ot\eps)\circ\Delta =\Id_H = (\eps\ot\Id_H)\circ\Delta\ , & \label{counitalDelta}
\eqaend
where $\eps:H\to \complex$ is the counit and $\Delta : H\rightarrow H
\ot H$ is the coproduct. The relation \eqref{coassoc} means that the
coalgebra is \emph{coassociative}. Throughout we use the usual
Sweedler notation $\Delta(h)=\sum\, {h_{(1)}\ot h_{(2)}}$ with
$h,h_{(1)},h_{(2)} \in H$
and suppress the summation.

A \emph{quasi-triangular bialgebra} is a pair $(H,\mcR)$ where $H$ is a
bialgebra, and $\mcR = \mcR_{(1)} \ot \mcR_{(2)} \in H \ot H$ is an
invertible element which obeys
\be
( \Delta \ot \Id_H ) (\mcR) = \mcR_{13}\, \mcR_{23} \ , \qquad (
\Id_H \ot \Delta ) (\mcR) = \mcR_{13} \, \mcR_{12} \ , \qquad ( \tau \circ \Delta ) (h) = \mcR \, \Delta(h) \, \mcR^{-1} \label{quasitriangular}
\ee
for all $h \in H$, where $\mcR_{12} = \mcR_{(1)} \ot \mcR_{(2)} \ot
1_H$, $\mcR_{13} = \mcR_{(1)} \ot 1_H \ot \mcR_{(2)}$, $\mcR_{23} =
1_H \ot \mcR_{(1)} \ot \mcR_{(2)}$ with $1_H$ the unit of $H$, and we
abbreviate the product map $\mu :H\ot H \to H$ by $\mu ( h \ot h'\, ) =h \, h'$ for all $h, h' \in H$. Here we have defined the \emph{transposition map} $\tau : H \ot H \to H \ot H$ as
\be
\tau ( h \ot h'\, ) :=h' \ot h  \label{tau}
\ee
for all $h, h' \in H$. 

Through the transposition map the \emph{co-opposite coproduct} $\Delta^{\mbox{\scr{op}}}: H \to H \ot H$ is defined by 
\be 
\Delta^{\mbox{\scr{op}}}(h):=( \tau \circ \Delta ) (h)= h_{(2)} \ot h_{(1)}  \ .
\ee
Then $H$ is a cocommutative coalgebra if $\Delta^{\mbox{\scr{op}}}(h) = \Delta(h)$ for all $h \in H$. If $(H,\mcR)$ is a quasi-triangular bialgebra then cocommutativity simply means that $\Delta(h) \, \mcR = \mcR\, \Delta (h)$ for all $h \in H$ as can be easily seen from \eqref{quasitriangular}; in general the element $\mcR$ intertwines the action of the coproduct $\Delta$ with the co-opposite coproduct $\Delta^{\mbox{\scr{op}}}$. 

 A \emph{Hopf algebra} over $\complex$ is a bialgebra $H$ equipped with an algebra anti-automorphism $S: H\to H$ called the antipode satisfying
\beq
\mu \circ (\Id_H\ot S)\circ \Delta=\eta_{{1_H}}\circ\eps=\mu\circ(S\ot \Id_H)\circ \Delta \ , \label{antipode}
\eeq
where $\eta_{{h}}:\complex \to H$ is the unit homomorphism with $\eta_{{h}}(1)=h$ for $h\in H$. A quasi-triangular Hopf algebra $(H, \mcR)$ consists of a quasi-triangular structure $\mcR$ on the underlying bialgebra of~$H$.

In this paper we will be primarily interested in the large class of Hopf algebras $H$ which arise as universal enveloping algebras $U(\mfg)$ of Lie algebras $\mfg$. The algebra $U(\mfg)$ is constructed by taking the quotient of the tensor algebra $T(\mfg)=\bigoplus_{k \geq 0}\, \mfg^{\ot k} =\complex \oplus \mfg \oplus (\mfg\ot\mfg) \oplus \cdots$ by the two-sided ideal $\mc I$ generated by elements of the form $x\ot y - y\ot x -[x,y]$, where $x,y\in \mfg$. Next we equip $U(\mfg)$ with the symmetric coalgebra structure 
\be
\begin{split}
& \Delta \,:\, U(\mfg) \ \longrightarrow \ U(\mfg)\ot U(\mfg) 
\ , \qquad \Delta(x)=x \ot 1 +1 \ot x \ , \\[4pt]
& \eps \,:\, U(\mfg) \ \longrightarrow \ \complex \qquad\qquad\quad  ,
\qquad \eps(x)=0 \ , \\[4pt]
& S \,:\, U(\mfg) \ \longrightarrow \ U(\mfg) \qquad\quad\   , \qquad S(x) =-x \label{costructure}
\end{split}
\ee
defined on primitive elements $x \in \mfg$ and extended to all of
$U(\mfg)$ as algebra (anti-)homomorphisms. The desired Hopf algebra $H$ is then $U(\mfg)=T(\mfg)/\mc I$ with these structure maps. Finally, we further equip $H$ with the trivial quasi-triangular structure 
\be
\mcR_0 = 1\ot 1 \ , \qquad 1_H := 1 \label{R0}
\ee 
which turns it into a cocommutative quasi-triangular Hopf algebra.

A Hopf algebra $H$ can act on a complex vector space $V$ to give a
representation of $H$ on $V$. In particular, a \emph{left action} of $H$ on $V$ is a pair ($\lambda , V$),
where $\lambda :H\ot V \to V$ is a linear map, $\lambda (h\ot v)
=:\lambda_h(v)$, such that $\lambda_{h\, g}(v)=\lambda_h(\lambda_g
(v))$ and $\lambda (1_H\ot v)=v$, where $g,h\in H$ and $v\in V$. It is
customary to denote the action of $H$ on $V$ by $\tri$ and write the
above relations as
\be
h \tri v \ \in \ V \ , \qquad (h\, g) \tri v =h\tri(g\tri v) \ , \qquad 1_H\tri v =v \ .
\ee
Such a vector space is called a \emph{left $H$-module}. If $V$ carries additional structure, for example if it is an algebra $(A, \mu_A)$ where $\mu_A : A\ot A\to A$ is the product on $A$, or a coalgebra $(C, \Delta_C)$ where  $\Delta_C : C\rightarrow C \ot C$ is the coproduct on $C$, then we demand that the action of $H$ is covariant in the sense that it preserves the additional structure of $V$. Thus we say that a unital algebra $(A, \mu_A)$ over $\complex$ is a \emph{left $H$-module algebra} if $A$ is a left $H$-module and
\be
h\tri (a\, b) = h \tri \mu_A (a\ot b) = \mu_A \big(\Delta (h) \tri (a
\ot b) \big)= \big(h_{(1)} \tri a\big) \, \big(h_{(2)} \tri b\big) \ , \label{tri}
\ee
where $h\in H$ and $a,b \in A$. Here we abbreviate $\mu_A (a \ot
b)=a\, b$; we use this notation throughout when no confusion arises. Likewise, a counital coalgebra $(C, \Delta_C)$ is a \emph{left H-module coalgebra} if $C$ is a left $H$-module and
\be
 ( h\tri c)_{(1)}\ot (h\tri c)_{(2)}  = \Delta_C (h\tri c) = \Delta(h) \tri \Delta_C(c)= \big( h_{(1)}\tri c_{(1)}\big)\ot \big(h_{(2)} \tri c_{(2)}\big) \ ,
\ee
where $h\in H$ and $c\in C$. Similar definitions hold for right $H$-modules, and right $H$-module algebras and coalgebras.

From the above definitions it is understood that a left or right $H$-module is a representation by an algebra $A$ or a coalgebra $C$ of the Hopf algebra $H$; if $H$ is modified then so is the representation. Such modifications were introduced by Drinfel'd. Let $H[[\hbar]]$ denote the $\hbar$-adic completion of $H$ consisting of all formal $H$-valued power series in a deformation parameter $\hbar$. A \emph{Drinfel'd twist} is an invertible element $F = F_{(1)} \ot F_{(2)} \in H[[\hbar]]\ot H[[\hbar]]$ that satisfies the two conditions
\eqa
&(F\ot1_H)\, (\Delta\ot\Id_H)(F) = (1_H\ot F)\, (\Id_H\ot\Delta)(F) \ , & \label{2cocycle}\\[4pt]
&\ \  (\eps\ot\Id_H)(F) =1_{H}=  (\Id_H\ot\eps)( F) \ . & \label{counital}
\eqaend
We will further demand that $F=1_H \ot 1_H + \mc O (\hbar)$ is a deformation of the trivial twist, which is always formally invertible for sufficiently small $\hbar$. By these two conditions $F$ is a counital 2-cocycle which can be used to define a new Hopf algebra $H_F$ with the same underlying algebra as $H[[\hbar]]$ but with a twisted coalgebraic structure given by the twisted coproduct 
\beq
\Delta_F (h)= F\, \Delta (h)\, F^{-1} \label{deltaF}
\eeq
and the twisted antipode 
\beq
S_F (h) = U_F\, S(h)\, U_F^{-1} \qquad\trm{where}\quad U_F=\mu\circ (\Id_H\ot S)(F) \label{SF}
\eeq
for $h\in H$. This new bialgebra $H_F$ is called a \emph{twisted Hopf algebra}; coassociativity and counitality of \eqref{deltaF} follow respectively from the $2$-cocycle condition \eqref{2cocycle} and the counital condition \eqref{counital}. If $(H, \mcR)$ is a quasi-triangular Hopf algebra then the quasi-triangular structure is also twisted by the formula
\be 
\mcR_F:=\tau(F) \, \mcR \, F^{-1} = F_{21} \, \mcR \, F^{-1} \ , \label{RF}
\ee
where $F_{21}=F_{(2)} \ot F_{(1)}$ in Sweedler notation. The Hopf
algebra $H_F$ need not be cocommutative even if $H$ is cocommutative; this may be checked by calculating the twisted co-opposite coproduct 
\be 
\Delta _F^{\mbox{\scr{op}}} (h) = \mcR_F \, \Delta_F (h)\, \mcR_F^{-1} \ , \label{deltaFcop}
\ee 
where $h \in H$.

For the twisted Hopf algebra $H_F$ to act covariantly on a left $H$-module algebra $(A,\mu_A)$ we need to twist (deform) the binary product $\mu_A : A\ot A\to A$ to a new product defined by
\be 
a\star b = \mu_A \big( F^{-1}\tri (a\ot b)\big)  = \big( F_{(1)}^{-1}\tri a\big) \,\big(F_{(2)}^{-1} \tri b \big)\ , \label{star}
\ee
for $a,b \in A$. The deformed product is called a \emph{star product}
and $(A[[\hbar]],\star)$ is a deformation quantization of $(A,
\mu_A)$; indeed one has $a \star b = a\, b+\mc{O}(\hbar)$. The twist cocycle condition \eqref{2cocycle} ensures associativity of the star product \eqref{star}, while the counital condition \eqref{counital} implies that if $(A,\mu_A )$ is unital with unit $1_A$ then $(A[[\hbar]],\star)$ is also unital with the same unit.


\subsection{Quasi-Hopf algebras and cochain twist quantization \label{qha}}

In this paper we are concerned with nonassociative twist deformations,
therefore we will be using an appropriate generalisation of a Hopf
algebra, called a quasi-Hopf algebra~\cite{Dri89}. To explain what a
quasi-Hopf algebra is let us begin by defining the notion of a
\emph{quasi-bialgebra}. This is simply a bialgebra $H$ where
coassociativity is required to hold only up to a $3$-cocycle $\phi$,
i.e. the condition \eqref{coassoc} is substituted by 
\be
(\Id_H\ot\Delta) \circ\Delta (h) =
\phi\,\big[(\Delta\ot\Id_H)\circ\Delta (h) \big] \, \phi^{-1} \ , \label{qcoassoc}
\ee
where $h \in H$ and $\phi =\phi_{(1)} \ot \phi_{(2)} \ot \phi_{(3)} \in H \ot H \ot H$ is an invertible $3$-cocycle (see e.g.~\cite{Majid:book}) in the sense that
\be
(1_H \ot \phi)\, \big[(\Id_H \ot \Delta \ot \Id_H)(\phi) \big]\, (\phi \ot 1_H) =\big[( \Id_H \ot \Id_H \ot \Delta )(\phi) \big]\,\big[(\Delta \ot \Id_H \ot \Id_H )(\phi) \big] \ . \label{3cocycle}
\ee
We say that $\phi$ is counital if it additionally satisfies the
condition
\be
(\eps \ot \Id_H \ot \Id_H )(\phi) = (\Id_H \ot \eps \ot \Id_H )(\phi) = (\Id_H \ot \Id_H \ot \eps)(\phi)=1_H\ot 1_H \ .
\ee
These two conditions on $\phi$ ensure that all distinct orderings of higher
coproducts by insertions of $\phi$ yield the same result and are
consistent with the counital condition \eqref{counitalDelta}. 

The definition of a quasi-triangular quasi-bialgebra is that of a quasi-triangular bialgebra with the first two axioms of \eqref{quasitriangular} modified by $\phi$ to 
\be
(\Delta \ot \Id_H ) (\mcR) = \phi_{321} \, \mcR_{13} \,
\phi_{132}^{-1} \, \mcR_{23} \, \phi \ , \qquad (\Id_H \ot \Delta ) (\mcR) = \phi_{231}^{-1} \, \mcR_{13} \, \phi_{213} \, \mcR_{12} \, \phi^{-1} 
\ee
in the notation of Section~\ref{ha} with $\phi_{abc}:= \phi_{(a)} \ot \phi_{(b)} \ot \phi_{(c)}$, while the third axiom of \eqref{quasitriangular} remains unchanged. 

A \emph{quasi-Hopf algebra} $\mcH =(H,\phi)$ is a quasi-bialgebra $H$
equipped with an antipode that consists of two elements $\alpha , \beta \in H$ and an algebra anti-automorphism $S:H\to H$ obeying
\eqa
S(h_{(1)}) \, \alpha \, h_{(2)}=\eps(h) \, \alpha \ , & \qquad h_{(1)} \, \beta \, S (h_{(2)})=\eps(h)\,\beta \ , \label{qantipode1}\\[4pt]
\phi_{(1)}\, \beta \, S(\phi_{(2)}) \, \alpha \, \phi_{(3)} = 1_H \ , & \qquad S\big(\phi^{-1}_{(1)}\big) \, \alpha \, \phi^{-1}_{(2)} \, \beta \, S \big(\phi^{-1}_{(3)}\big) =1_H \ ,  \label{qantipode2}
\eqaend
for all $h\in H$. The antipode is determined uniquely only up to the transformations
\be 
S'(h)=u\,S(h)\,u^{-1} \ , \qquad \alpha'=u\,\alpha \ , \qquad \beta'=\beta \, u^{-1} \ , \label{transformation}
\ee
for any invertible element $u\in H$ and any $h\in H$. When $\phi=1_H
\ot 1_H \ot 1_H$ is the trivial 3-cocycle, the conditions \eqref{qantipode2} imply that $\alpha\,\beta=\beta\,\alpha =1_H$, and the symmetry \eqref{transformation} allows us to suppose without loss of generality that $\alpha=\beta=1_H$. Then \eqref{qcoassoc} reduces to the coassociativity condition \eqref{coassoc} and \eqref{qantipode1} to the usual definition of an antipode given by \eqref{antipode}, thus the quasi-Hopf algebra $\mcH$ becomes a coassociative Hopf algebra. 

A useful way to construct a quasi-Hopf algebra is to start with a Hopf
algebra $H$ and an invertible twist element $F\in H[[\hbar]]\ot
H[[\hbar]]$ that does not satisfy the cocycle condition \eqref{2cocycle}. In particular, if $(H,\phi,\mcR)$ is a quasi-triangular quasi-Hopf algebra and $F$ is an arbitrary invertible element in $H[[\hbar]] \ot H [[\hbar]]$ obeying \eqref{counital}, then $(H_F, \phi_F, \mcR_F)$ defined as follows is also a quasi-triangular quasi-Hopf algebra. It has the same algebra and counit as $H$, with twisted coproduct and quasi-triangular structure defined by the same formulas \eqref{deltaF} and \eqref{RF}, with twisted antipode
\be
S_F = S \ , \qquad \alpha_F=S\big(F^{-1}_{(1)}\big)\, \alpha \, F^{-1}_{(2)} \ , \qquad \beta_F = F_{(1)} \, \beta \, S\big(F_{(2)}\big) \ , 
\ee
and with twisted $3$-cocycle given by the coboundary
\be
\phi_F =\p^* F:= F_{23}\, \big[(\Id_H\ot\Delta)(F)
\big]\,\phi\,\big[(\Delta\ot\Id_H )\big( F^{-1}\big) \big] \, F_{12}^{-1} \ ,  \label{coboundary} \\
\ee
where $F_{23}=1_H\ot F$, $F_{12}^{-1}=F^{-1}\ot1_H$ and $\phi_F \in
H[[\hbar]]\ot H[[\hbar]]\ot H[[\hbar]]$ is called the
\emph{associator} (see e.g.~\cite{Majid:2005}). A Hopf algebra $H$
viewed as a trivial quasi-Hopf algebra has $\alpha = \beta =1_H$ and
the symmetry \eqref{transformation}. Twisting $H$ with the counital
$2$-cochain twist $F$ then provides a quasi-Hopf algebra $\mcH_F =
(H_F,\phi_F)$ with $S_F = S$, $\alpha_F = \mu \circ (S \ot \Id_H) (
F^{-1})$ and $\beta_F = \mu \circ ( \Id_H \ot S) (F) = \alpha_F^{-1}$ which by \eqref{transformation} is equivalent to \eqref{SF}. The twisted coproduct $\Delta_F$ fails to satisfy \eqref{coassoc}, and in particular \eqref{qcoassoc} is a consequence of this definition. 

A left $H$-module algebra $(A, \mu_A)$ is then twisted to a
\emph{nonassociative} algebra $(A[[\hbar]], \star)$ by the same
formula \eqref{star} with the associator appearing when we rebracket
products of three elements as
\be
(a\star b)\star c =\big(\phi_{F\,(1)}\tri a\big)\star \big[\big(\phi_{F\,(2)}\tri b\big)\star\big(\phi_{F\,(3)}\tri c\big)\big] \ , \label{product of 3}
\ee
for $a,b,c \in A$. The cocycle condition (\ref{3cocycle}) on $\phi_F$
ensures that the distinct ways of rebracketing higher order products by
inserting $\phi_F$ all yield the same result.


\subsection{Twist quantization functor \label{categories}}

A natural way to deal with both noncommutative and nonassociative
structures arising as above
is through the formalism of braided monoidal categories. The algebras
encountered above are ``braided-commutative'' and
``quasi-associative'', in the sense that they are noncommutative and
nonassociative but in a controlled way by means of a braiding and a multiplicative
associator, respectively. This means that the algebras
\emph{are} commutative and associative when regarded as objects of a suitable braided monoidal
category which is different from the usual category of complex vector
spaces. The twist deformation
quantization described above can then be regarded as a functor that
yields algebras in such a braided monoidal category, and at the same
time quantises all other covariant structures with respect to a
symmetry. We briefly review this framework here as we will make
reference to it later on, and because it connects with some of the
constructions of~\cite{Mylonas:2012pg} as 
explained in Section~\ref{intro}. 

A monoidal category $\CCC$ consists of a collection of objects
$V,W,Z,\dots$ with a tensor product between any two objects and a
natural associativity isomorphism $\Phi_{V,W,Z}:(V\otimes W)\otimes Z\to
V\otimes(W\otimes Z)$ for any three objects obeying the pentagon
identity, which states that the two ways of rebracketing morphisms
$\big((U\otimes V)\otimes W\big)\otimes Z\to
U\otimes\big(V\otimes(W\otimes Z)\big)$ are the same. Then MacLane's
coherence theorem states that all different ways of inserting
associators $\Phi$ as needed to make sense of higher order rebracketed
expressions yield the
same result. A braiding on $\CCC$ is a natural commutativity
isomorphism $\Psi_{V,W}:V\otimes W\to W\otimes V$ for any pair of
objects which is
compatible with the associativity structure in a natural way. If
$\CCC$ is the category of complex vector spaces, then the associator $\Phi$
is the identity morphism and the braiding $\Psi$ is the transposition morphism.

If $\hil=(H,\phi)$ is a quasi-Hopf algebra, we take $\CCC$ to be the
category ${}_H\CCM$ of left $H$-modules. This is a monoidal category
with tensor product defined via the coproduct $\Delta$ and with
associator given by
\beq
\Phi_{V,W,Z}\big((v\otimes w)\otimes z\big) = \big(\phi_{(1)}\triangleright
v\big)\otimes \big[\big(\phi_{(2)}\triangleright w\big)\otimes
\big(\phi_{(3)}\triangleright z\big)\big]
\eeq
for all $v\in V$, $w\in W$ and $z\in Z$. If in addition $H$ is
quasi-triangular then there is a braiding defined by
\beq
\Psi_{V,W}(v\otimes w) = \big(\mcR_{(1)}\triangleright
w\big)\otimes\big(\mcR_{(2)}\triangleright v\big)
\eeq
for all $v\in V$ and $w\in W$.

Given a cochain twist $F\in H[[\hbar]]\otimes H[[\hbar]]$, the
constructions of this section determine a functorial isomorphism of braided monoidal
categories
\beq
\CCF_F\,:\, {}_H\CCM \ \longrightarrow \ {}_{H_F}\CCM
\eeq
which acts as the identity on objects and morphisms, but intertwines
the tensor, braiding and associativity structures. In particular, the
covariance condition (\ref{tri}) means that the product map $\mu_A$ is
a morphism in the category ${}_H\CCM$; hence $\CCF_F$ functorially
deforms $H$-module algebras into $H_F$-module algebras, and in this sense
it may be regarded as a ``twist quantization functor''.

In our main
case of interest in this paper, we will take $H=U(\mfg)$ to be the
universal enveloping algebra of a Lie algebra $\mfg$ of symmetries
acting on a manifold $\cal M$; then the Hopf algebra $H$ acts on the algebra of smooth
functions $A=C^\infty({\cal M})$, and by functoriality of the twist
deformation it is also quantized to a generically noncommutative and
nonassociative algebra $A_F$, which is in fact commutative and
associative in the category ${}_{H_F}\CCM$. Similarly, the 
exterior algebra of differential forms $\Omega^\bullet({\cal M})$ is
quantized to $\Omega_F^\bullet({\cal M})$ as a differential calculus
on $A_F$; in this way any geometry can be systematically
quantized with respect to a symmetry and a choice of
2-cochain~$F$.


\newsection{Cochain twist quantization of parabolic $R$-flux backgrounds \label{deformation}}

In this section we employ the formalism of Section~\ref{td} to study
nonassociative deformations of certain non-geometric closed string backgrounds. In order to make contact with the key ideas
of~\cite{Mylonas:2012pg} we will initially study standard deformation
quantization on the cotangent bundle of a closed string vacuum
and subsequently add a constant background $R$-flux. This approach has
the advantage of illuminating pertinent non-local and non-geometric
symmetry transformations analogous to the ones which arise on T-folds
induced by parabolic monodromies.


\subsection{Quantum phase space \label{qps}}

Let us begin by considering a manifold $M$ of dimension $d$ with
trivial cotangent bundle ${\cal M} := T^*M \cong M\times
(\real^d)^*$
and coordinates $x^I = (x^i, p_i)$, where $I=1, \dots,2d$, $(x^i) \in
M$, $(p_i) \in (\real^d)^*$ and $i=1, \dots,d$. Throughout we use
upper case indices for the full phase space while lower case indices
will be reserved for position or momentum space individually. Consider
the abelian Lie algebra $\mfh = \real^d \oplus ( \real^d )^*$ of
dimension $2d$ generated by $P_i$ and $\tP^i$. It is realised on
${\cal M}$ by its action on the algebra of smooth complex functions $C^\infty({\cal M})$ which we take to be given by the vector fields 
\be
P_i\tri f := \p_i f \qquad \mbox{and}\qquad \tP^i \tri f :=  \tilde\p^i f \ , \label{rep}
\ee
where $f\in C^\infty({\cal M})$, $\p_i=\frac{\p}{\p x^i}$ and
$\tilde\p^i=\frac{\p}{\p p_i}$. For constant vectors $a=(a^i)\in
\real^d$ and $a=( \tilde a_i) \in (\real^d)^*$ we can define $P_a=a^i \, P_i$ and $\tP_{\tilde a} = \tilde a_i \, \tP^i$ which as vector fields on ${\cal M}$ translate $x^i$ and $p_i$ by $a^i$ and $\tilde a_i$ respectively, hence $\mfh$ is the classical phase space translation algebra on $M$.

We can now construct the related Hopf algebra $K$ that acts on an algebra $(A, \mu_A)$ in the usual way, i.e. we consider the universal enveloping algebra $K=U(\mfh)$ and equip it with the coalgebra structure from Section~\ref{ha}. In particular, the action of $K$ on $A=C^\infty({\cal M})$ is given by \eqref{rep} extended covariantly to all elements in $K$ using linearity and the Leibniz rule for the vector fields $\p_i$ and $\tilde\p^i$.

Phase space quantization is carried out simply by twisting $K$ in the manner described in Section~\ref{ha}. A suitable abelian twist $F\in K[[\hbar]] \ot K [[\hbar]]$ is given by
\be
F=\exp \big[ \mbox{$-\frac{\ii \hbar}{2}$}\, (P_i \ot \tP^i -\tP^i \ot P_i ) \big] \ , \label{F}
\ee
where $\hbar$ is the deformation parameter. In this simple case
$\Delta_F=\Delta$, where the coproduct $\Delta$ is defined in \eqref{costructure}, and thus the twisted Hopf algebra $K_F$ is cocommutative. The twisted quasi-triangular structure is easily calculated from \eqref{RF} and is given by
\be 
\mc Q = F^{-2} =  \exp \big[ \ii \hbar\, (P_i \ot \tP^i -\tP^i \ot P_i ) \big] \ . \label{Q}
\ee

We may now deform any left (or right) $K$-module algebra $(A, \mu_A)$
using \eqref{star} and the relevant action. Let us do this for the
algebra of functions on ${\cal M}$, i.e. we set $A=C^\infty({\cal M})$
and $\mu_A (f\ot g) = f\, g$ the pointwise multiplication of functions, and derive its deformation quantization $(C^\infty({\cal M})[[\hbar]], \moyal )$; the star product given by \eqref{star} is 
\be
f \moyal g =\mu_A \Big(\exp{\big[ \mbox{$\frac{\ii\hbar}{2}$}\,\big(\p_i \otimes \tilde\p^i - \tilde\p^i \otimes \p_i \big) \big]}(f \otimes g ) \Big)\ , \label{moyal}
\ee
where the action \eqref{rep} has been used. This \emph{noncommutative}
star product is the canonical associative Moyal-Weyl star product familiar
from quantum mechanics. With its use, the usual quantum phase space commutation relations are calculated as
\be 
[x^i, p_j ]_\moyal = \ii \hbar \, \delta^i{}_j \ , \qquad [x^i, x^j ]_\moyal =0= [p_i, p_j ]_\moyal \ ,
\ee
where $[f,g]_\moyal :=f \moyal g -g \moyal f $ for all $f,g\in C^\infty({\cal M})$.


\subsection{Noncommutative quantum phase space \label{ncqps}}

Let us endow $M$ with a constant trivector $R=\frac {1}{3!}\, R^{ijk}\, \p_i \wg \p_j \wg \p_k$ which is T-dual to the background $H$-flux of a non-trivial closed string $B$-field. To bring $R$ into the twist quantization scheme we introduce a family of antisymmetric linear combinations of the generators of $\mfh$ as
\be 
\tM_{ij} :=M^{\tp}_{ij}= \tp_i \, P_j - \tp_j \, P_i \ ,
\ee 
which we will regard as parametrized by constant momentum surfaces
$\tp =(\tp_i) \in (\real^d)^*$. The generators $\tM_{ij}$ are unique
in the sense that they are the only rank two tensors constructed by
primitive elements in $\mfh$ to lowest order that can be non-trivially contracted with the constant antisymmetric trivector components $R^{ijk}$. The restriction to constant momentum surfaces here will ensure (co)associativity, but will be relaxed in the next subsection.

The unique twist element that can be constructed in this way from generators of $\mfh$ is the abelian twist $\tF_R\in K[[\hbar]] \ot K[[\hbar]]$ given by
\be
\tF_R =\exp{\big( \mbox{$-\frac{\ii\hbar}{8}$}\,R^{ijk}\, (\tM_{ij} \ot P_k -  P_i \ot \tM_{jk}) \big)} \ , \label{tFR}
\ee
where $\hbar$ is the deformation parameter. The Hopf algebra $K$ is twisted to a new Hopf algebra $K_{\tF_R}$ which is cocommutative with quasi-triangular structure given by 
\be 
\tmcR = \tF_R{}^{-2} \ ,
\ee
and the algebra of functions $( C^\infty({\cal M}), \mu_A )$ is quantized to $( C^\infty({\cal M})[[\hbar]], \tAs :=\As_\tp )$. The star product $\tAs$ is calculated by \eqref{star} as
\be
f \tAs g =\mu_A \Big(\exp{\big[ \mbox{$\frac{\ii\hbar}{2}$}\, R^{ijk}\, \tp_k \,  \p_i \otimes \p_j \big]}(f \otimes g ) \Big) \label{tAs}
\ee
for all $f,g \in C^\infty({\cal M})$, and it is a \emph{family of Moyal-Weyl products}. This can be seen by calculating the $\tAs$-commutators $[f,g]_\tAs :=f \tAs g - g \tAs f$ on phase space coordinate functions where we find
\be 
[x^i,x^j]_\tAs = \ii \hbar \, \theta^{ij}(\tp)  \ , \qquad [x^i, p_j ]_\tAs =0= [p_i, p_j ]_\tAs  \ ,
\ee
where 
\be 
\theta^{ij}(\tp) = R^{ijk}\,\tp_k \ . \label{theta}
\ee 
This reveals that upon twisting, the cotangent bundle ${\cal M}=T^*M$
is deformed to $
M_{\theta(\tp)} \times (\real^d)^*$; hence configuration space is
quantized to a family of noncommutative spaces $M_{\theta(\tp)}$ with
constant noncommutativity parameters proportional to the constant
background $R$-flux and parametrized by the surfaces of constant
momentum $\tp=(\tp_i)\in (\real^d)^*$. This is similar to what happens
in the associative $Q$-flux T-duality frame with parabolic monodromy, where the configuration
space noncommutativity is proportional to the winding numbers $w\in\zed^d$ of
closed strings, i.e. $[x^i, x^j] = \ii \hbar\, Q^{ij}{}_k \, w^k$.

The above construction is further extended to give a noncommutative quantum phase space if we use the abelian twist 
\be 
\tmcF := \tF_R \, F =F \, \tF_R  = \exp{\big[ \mbox{$-\frac{\ii\hbar}{2}$}\, \big(\mbox{$\frac 14$}\,R^{ijk}\, (\tM_{ij} \ot P_k - P_i \ot \tM_{jk}) + P_i \ot \tilde P^i - \tilde P^i \ot P_i\big)\big]} \ . \label{tmcF}
\ee
The quasi-triangular Hopf algebra $(K , \mcR_0)$ is twisted to the cocommutative quasi-triangular Hopf algebra $(K_\tmcF ,\bar{\mc Q})$ with quasi-triangular structure 
\be 
\bar{\mc Q}= \tmcF^{-2} = \mc Q \, \tmcR = \tmcR \,  \mc Q \ ,
\ee
and quantization of $( C^\infty({\cal M}), \mu_A )$ is given by $(C^\infty({\cal M})[[\hbar]], \ts :=\star_\tp )$, where 
\be
f \ts g =\mu_A \Big(\exp{\big[ \mbox{$\frac{\ii\hbar}{2}$}\,\big( R^{ijk}\, \tp_k \,  \p_i \otimes \p_j + \p_i \otimes \tilde\p^i - \tilde\p^i \otimes \p_i \big)\big]} (f \otimes g ) \Big)\ , \label{ts}
\ee
for all $f,g \in C^\infty({\cal M})$. In this case the full cotangent bundle ${\cal M}$ becomes a noncommutative quantum phase space with commutation relations 
\be
[x^i,x^j]_\ts = \ii \hbar \, \theta^{ij}(\tp) \qquad , \qquad [x^i,p_j]_\ts = \ii \hbar \, \delta^i{}_j \qquad , \qquad [p_i,p_j]_\ts = 0
\ee
where $[f,g]_\ts :=f \ts g - g \ts f$ for all $f,g \in C^\infty({\cal
  M})$. In particular, the zero momentum surface $\tp=0$ recovers the
canonical Moyal-Weyl product $\moyal=\star_0$ on phase space from Section~\ref{qps}.

The star product \eqref{ts} first appeared in~\cite{Mylonas:2012pg} as
the restriction of a nonassociative star product to slices of constant
momentum in phase space. However, in the context of cocycle twist
quantization its origin can be traced back to the unique choice of
contraction of the $R$-flux $R^{ijk}$ which is compatible with the
``minimal" (translation) symmetries of ${\cal M}$ and
associativity. We will see below how this choice
naturally extends to dynamical momentum and leads to the
nonassociative star product of~\cite{Mylonas:2012pg}, thus further
clarifying the operations of restricting to constant momentum and of reinstating dynamical momentum dependence that were used for the derivation of the quantized associator in~\cite{Mylonas:2012pg}.


\subsection{Let's twist again \label{lta}}

As discussed in~\cite{Mylonas:2012pg}, the trivector $R$ has no
natural geometric interpretation on configuration space except via
T-duality. On the other hand, it is a $3$-form on phase space ${\cal
  M}$ which is in fact the curvature of a non-flat $U(1)$ gerbe in momentum
space. The presence of this $3$-form enhances the symmetries of ${\cal
  M}=T^*M$ and thus the abelian Lie algebra $\mfh$ should be enlarged in
order to accommodate the new symmetries. For this, we extend $\mfh$ to
the non-abelian nilpotent Lie algebra $\mfg$ of dimension $\frac1{2}\,d\, (d+3)$ generated by $P_i$, $\tP^i$ and $M_{ij} = -M_{ji}$ with commutation relations given by 
\be
[\tilde P^i , M_{jk}]= \delta^i{}_{j} \, P_k - \delta^i{}_{k} \, P_j \ , \label{cr}
\ee
while all other commutators are equal to zero. The actions of $P_i$ and $\tP^i$ on $C^\infty({\cal M})$ are still given by \eqref{rep} from which we find the action of $M_{ij}$ on $C^\infty({\cal M})$ to be
\be
M_{ij}\tri f := p_i \, \p_j f -p_j \, \p_i f \ \label{repM}
\ee
for all $f \in C^\infty({\cal M})$. Introducing elements $M_\s =\frac
12 \, \s^{ij}\, M_{ij} \in \mfg$, where $\s^{ij}=-\s^{ji} \in \real$, we see that $M_{\s}$ generates the non-local coordinate transformations
\be
x^i \ \longmapsto \ x^i + \s^{ij}\, p_j \ , \qquad p_i \ \longmapsto \
p_i 
\label{Bopp}\ee
that mix positions and momenta, which in quantum mechanics are called
\emph{Bopp shifts}. This symmetry is reminscent of those encountered in
T-folds ($Q$-space), where diffeomorphism symmetries include T-duality
transformations that mix positions with winding
numbers which are T-dual to the conjugate momenta. This points to the use of doubled
geometry,\footnote{It is possible to extend our
  star products below to T-duality covariant star products defined on
  double phase space, as in~\cite{Bakas:2013}; a field theory written
  in this formalism is manifestly $O(d,d)$-invariant. However,
  in order to avoid overly cumbersome equations with essentially the
  same generic features, for simplicity we
  write all formulas below only in the $R$-flux duality frame.} while
here we are on a phase space of $2d$ coordinates. In this sense the
symmetries (\ref{Bopp}) can be regarded as the analog of T-duality transformations in our algebraic framework.

One may now proceed as we did previously to find the related quasi-triangular Hopf algebra $(H, \mcR_0)=(U(\mfg), \Delta, \eps, S, \mcR_0)$, i.e. we equip the universal enveloping algebra $U(\mfg)$ with the structure maps \eqref{costructure} and the quasi-triangular structure \eqref{R0}. Then $H$ can be twisted by the abelian Drinfel'd twist
\be
F_R =\exp{\big( \mbox{$-\frac{\ii\hbar}{8}$}\,R^{ijk}\, (M_{ij} \ot P_k - P_i \ot M_{jk} )\big)} \ . \label{FR}
\ee
The result is a cocommutative twisted Hopf algebra $H_{F_R}$ with
quasi-triangular structure $\mcR=F_R{}^{-2}$. We can use $H_{F_R}$ to
twist the algebra of functions $(C^\infty({\cal M}), \mu_A)$, and the resulting star product has the form
\be
f \As_p \, g := f \As g =\mu_A \Big(\exp{\big[ \mbox{$\frac{\ii\hbar}{2}$}\, R^{ijk}\, p_k \,  \p_i \otimes \p_j \big]}(f \otimes g ) \Big) \label{ast}
\ee
for all $f,g \in C^\infty({\cal M})$, and it is a noncommutative, associative Moyal-Weyl type star product similar to the one found in Section~\ref{ncqps}. The algebra $(C^\infty({\cal M}), \mu_A)$ is hence quantized to  $(C^\infty({\cal M})[[\hbar]], \As_p := \As)$ and ${\cal M}$ acquires a spatial noncommutativity since 
\be 
[x^i, x^j ]_\As = \ii \hbar \, \theta^{ij} (p) \ , \qquad [x^i, p_j ]_\As=0=[p_i, p_j ]_\As  \ ,
\ee
where $[f,g ]_\As := f\As g -g \As f$ for all $f,g \in C^\infty({\cal M})$ and $\theta^{ij}(p)$ is defined by \eqref{theta}.

We can incorporate quantum phase space in this description using the method of Section~\ref{ncqps}. The pertinent non-abelian twist $\mcF\in H[[\hbar]] \ot H[[\hbar]] $ is 
\be
\mcF := F_R \,F= F \, F_R = \exp{\big[ \mbox{$-\frac{\ii\hbar}{2}$}\, \big(\mbox{$\frac 14$}\,R^{ijk}\, (M_{ij} \ot P_k - P_i \ot M_{jk} ) + P_i \ot \tilde P^i - \tilde P^i \ot P_i\big)\big]} \ , \label{mcF}
\ee
where we have used the antisymmetry of $R^{ijk}$. Equivalently,
instead of \eqref{mcF} we can use the action of \eqref{tmcF} on
$C^\infty({\cal M})$ to write
\be
\mcF = [\tmcF]_{\tp \to p}
\ee
where the operation $[-]_{\tp \to p}$ denotes the change from constant
to dynamical momentum.\footnote{We can also restrict to constant
  momentum by taking a double scaling limit $\hbar\to0$,
  $R^{ijk}\to\infty$ with $\bar R^{ijk}:=\hbar\, R^{ijk}$ held
  constant. This limit is equivalent to the restriction
  $\tF_R=[F_R]_{p\to \tp}$ when acting on $C^\infty({\cal M})$.}

The twist $\mcF$ is an invertible counital 2-cochain, hence $H_\mcF$ defines a quasi-Hopf algebra $\mcH =(H_\mcF, \phi)$ where the associator $\phi=\phi_\mcF$ calculated from \eqref{coboundary} is
\be
\phi = \exp\big(\mbox{$\frac{\hbar^2}{2}$}\,R^{ijk}\, P_i \ot P_j \ot P_k \big) \ . \label{associator}
\ee
Its coproduct $\Delta_{\mcF} : H_\mcF \to H_\mcF \ot H_\mcF$ is given by \eqref{deltaF}; calculating this on the generating primitive elements we get 
\be
\begin{split}
\Delta_{\mcF}(P_i) &= \Delta (P_i) \ , \\[4pt]
\Delta_{\mcF}(\tilde P^i) &= \Delta (\tilde P^i) +
\mbox{$\frac{\ii\hbar}{2}$} \, R^{ijk}\, P_j \ot P_k \ ,\\[4pt]
\Delta_{\mcF}(M_{ij}) &= \Delta (M_{ij}) - \ii\hbar\, ( P_i \ot P_j - P_j \ot P_i ) \ . \label{Delta}
\end{split}
\ee
In particular, $\mcH$ is a non-cocommutative quasi-Hopf algebra with quasi-triangular structure $\underline{\mcR} = \mcF^{-2}$, as a straightforward calculation of the co-opposite coproduct $\Delta_{\mcF}^{\mbox{\scr{op}}}$ on primitive elements using \eqref{deltaFcop} reveals.

A left (or right) $H$-module algebra $(A, \mu_A)$ can now be deformed
using \eqref{star} and the relevant action. Let us do this for the
algebra of functions on ${\cal M}$, i.e. we set $A=C^\infty({\cal M})$
and $\mu_A (f\ot g)= f\, g$, and derive its deformation quantization $(C^\infty({\cal M})[[\hbar]], \star)$ with the star product given by \eqref{star}. We find
\be
f \star g =\mu_A \Big(\exp{\big[ \mbox{$\frac{\ii\hbar}{2}$}\,\big( R^{ijk}\, p_k \,  \p_i \otimes \p_j + \p_i \otimes \tilde\p^i - \tilde\p^i \otimes \p_i \big)\big]}(f \otimes g ) \Big)\ , \label{sp}
\ee
where the actions \eqref{rep} and \eqref{repM} have been used. This is a \emph{nonassociative star product} and hence $(C^\infty({\cal M})[[\hbar]], \star)$ is a nonassociative algebra, i.e. the product of three functions is associative only up to the associator \eqref{associator}. This is expressed by \eqref{product of 3} which in this case can be written in the more explicit form
\be
\begin{split}
( f\star g) \star h = f \star ( g \star h ) + \sum_{n=1}^\infty \,
\frac{1}{n!}\, \Big(\, \frac{\hbar^2}{2}\, \Big)^n \, R^{i_1 j_1 k_1}\cdots
  R^{i_n j_n k_n}\, ( \p_{i_1}\cdots \p_{i_n} f ) \ \star  \\
  \star \ \big( ( \p_{j_1}\cdots \p_{j_n} g ) \star ( \p_{k_1}\cdots \p_{k_n} h ) \big) \ . \label{Rassociator}
\end{split}
\ee
From \eqref{Delta} we find the (modified) Leibniz rules
\be
\begin{split}
\p_i(f \star g) &= (\p_i f) \star g + f \star (\p_i g) \ , \\[4pt]
\tilde \p^i(f \star g) &= (\tilde \p^i f) \star g + f \star (\tilde \p^i g) + \mbox{$\frac{\ii \hbar}{2}$}\, R^{ijk} \, (\p_j f) \star (\p_k g) \ , \label{Leibniz}
\end{split}
\ee
and in particular $\p_i$ is a derivation of the star product. These
relations simply reflect the fact that the nonassociative $R$-flux background
exhibits space translation symmetry, but not momentum
translation symmetry, due to the dynamical nonassociativity. The loss of translation invariance in momentum
space is related to the violation of Bopp shift symmetry through
\be
\Delta_{\mcF}(\tilde P^i)-\Delta(\tilde P^i) = -\mbox{$\frac14$}\,
R^{ijk}\, \big(\Delta_{\mcF}(M_{jk})-\Delta(M_{jk})\big) \ ,
\ee
indicating that the Bopp shift generators $M_{ij}$ are 
observables which detect effects of nonassociativity.

The nonassociative star product \eqref{sp} coincides with the one
previously proposed in~\cite{Mylonas:2012pg} for functions on ${\cal
  M}$. To see that
\eqref{Rassociator} coincides with the associator derived
in~\cite{Mylonas:2012pg}, it is enough to compute it on the position
space plane waves $U_{\tilde a}=\e^{\ii \tilde a_i\, x^i}$ where $\tilde
a=(\tilde a_i)\in(\real^d)^*$ are constant vectors. One easily finds
\be
\big(U_{\tilde a}\star U_{\tilde b}\big)\star U_{\tilde c} = \varphi_R\big(\tilde
a\,,\,\tilde b\,,\,\tilde c\big) \ U_{\tilde a}\star \big(U_{\tilde b}\star U_{\tilde
  c}\big) \ ,
\ee
where
\be
\varphi_R\big(\tilde
a\,,\,\tilde b\,,\,\tilde c\big):= \exp\big(-\mbox{$\frac{\ii\hbar^2}2$}\,
R^{ijk}\, \tilde
a_i\, \tilde b_j\,\tilde c_k\, \big)
\ee
is the 3-cocycle of the group of translations that was obtained
in~\cite{Mylonas:2012pg,Bakas:2013}. 

Using \eqref{sp} we can calculate the $\star$-commutation relations on the coordinate functions of ${\cal M}$. We find
\be 
[x^i, x^j]_\star =\ii \hbar \, \theta^{ij}(p) \ , \qquad [x^i ,p_j]_\star = \ii \hbar\, \delta^i{}_j \ , \qquad [p_i ,p_j]_\star =0 
\label{starcommrels}\ee
where $[f,g]_\star :=f\star g-g\star f$ for all $f,g \in
C^\infty({\cal M})$ and $i,j,k = 1,\dots,d$. Nonassociativity of
\eqref{sp} can also be seen through the failure of the Jacobi identity
for this $\star$-commutator, analogously to (\ref{jacobiator}). As
explained in~\cite{Mylonas:2012pg}, the antisymmetric bracket
$[-,-]_\star$ gives rise to a Lie $2$-algebra structure on the
function algebra $C^\infty({\cal M})$; while Lie algebraic structures
give rise to associative deformations, Lie $2$-algebras lead to
nonassociative deformations.


\subsection{Integral formulas}

For later use, let us note that
the star product (\ref{sp}) can be written in upper case index notation as
\be
f\star g = \mu_A \Big( \exp\big(\mbox{$\frac{\ii\hbar}{2}$} \, \Theta^{IJ} \,\p_I \ot \p_J\big)  (f \ot g) \Big) \ , \label{sstar}
\ee
where $\partial_I= \frac\partial{\partial x^I}$ and
\be
\Theta=\big(\Theta^{IJ} \big)=\begin{pmatrix} R^{ijk}\, p_k & \delta^i{}_j \\
  -\delta_i{}^j & 0 \end{pmatrix} \label{Theta}
\ee
with $I,J \in \{1,\dots,2d\}$. The formula \eqref{sstar} for the
nonassociative star product was first derived in~\cite{Mylonas:2012pg}
using Kontsevich's deformation quantization; it is formally identical to the Moyal-Weyl star product, but it is nonassociative and its derivation is non-trivial. In practise, for explicit
calculations it is useful to employ an integral representation of
the star product; a Fourier integral formula for this product was derived
in~\cite[Section~4.3]{Mylonas:2012pg}
and in~\cite[Section~4.2]{Bakas:2013}. A related but more useful formula can be
easily derived by expressing $g(x)$ in terms of its Fourier transform $\hat g(k)$, where
$k\in{\cal M}^*$; here ${\cal M}:=T^*M\cong
M\times(\real^d)^*\cong\real^d\times(\real^d)^*$ and ${\cal
  M}^*\cong(\real^d)^*\times \real^d$. Then the derivative
$\partial_J$ turns into multiplication by $\ii k_J$ and we can
interpret the exponential $\exp\big(\mbox{$-\frac{\hbar}{2}$} \,
\Theta^{IJ}\, k_J\,\p_I  \big)$ as a shift operator to rewrite \eqref{sstar} as
\be
(f\star g)(x) = \int_{\Mcal^*}\, \dd^{2d} k \ f\big(x- \mbox{$\frac\hbar 2
  $}\, \Theta k \big) \, \hat g(k) \, \e^{\ii k_I\,x^I } \ ,
\ee
where $(\Theta k)^I:=\Theta^{IJ}\, k_J$.
Using the inverse Fourier transformation, this becomes
\be
(f\star g)(x) = \frac 1{(2\pi)^{2d}}\, \int_{\Mcal^*}\, \dd^{2d} k \
\int_\Mcal\, \dd^{2d} x'\ f\big(x- \mbox{$\frac\hbar 2$}\, \Theta k\big) \,
g(x'\,) \, \e^{\ii k_I\, (x-x'\,)^I} \ .
\ee
Since the matrix $\Theta$ is unimodular, we can use the inverse matrix ($B$-field)
\be
\Theta ^{-1}=\big(\Theta^{-1}_{IJ} \big)=\begin{pmatrix} 0 & -\delta^i{}_j \\
  \delta_i{}^j & R^{ijk}\, p_k \end{pmatrix} \label{Theta-inverse}
\ee
to change variables to
$z=-\frac\hbar2\, \Theta k$ and $w=x'-x$, and in this way we finally obtain a nice
expression in terms of a
twisted convolution product
\be \label{intstar}
(f\star g)(x) = \Big(\,\frac 1{\pi\,\hbar}\,\Big)^{2d}\,
\int_{\Mcal}\, \dd^{2d} z \ \int_{\Mcal}\, \dd^{2d} w\  f(x +z) \,
g(x+w) \, \e^{-\frac{2\ii}{\hbar}\,
  z^I\, \Theta^{-1}_{IJ}\, w^J}
\ee
that is often more convenient for computations than \eqref{sstar}; it
is also well-defined as a nonperturbative formula on the larger class of Schwartz functions on
phase space $\Mcal$.
One should not forget that the matrix $\Theta^{-1}$ in \eqref{intstar}
depends on $p_i$ and hence on $x^I$, but otherwise the expression is
again formally identical to the standard twisted convolution product formula for the
Moyal-Weyl product (see e.g.~\cite{Szabo:2001kg} for a review).

There are also integral formulas available for our twists. 
Changing the sign of $\Theta$ and dropping the multiplication operator
$\mu_A$ in \eqref{sstar}, we can similarly derive an integral formula
for the action of the twist on a pair of functions $f$ and $g$ (evaluated at $x$
and $y$ respectively) given by
\bea
&& \big( \mathcal F \triangleright (f \ot g) \big) (x,y) \\ && \qquad \qquad
= \Big(\,\frac 1{\pi\,\hbar} \,\Big)^{2d}\,
\int_{\Mcal} \, \dd^{2d} z \ \int_{\Mcal} \,
\dd^{2d} w \ \big(\e^{\frac{\ii}{\hbar}\,
  z^I\, \Theta^{-1}_{IJ}\, w^J} \,f\big) \ot \big(\e^{\frac{\ii}{\hbar}\,
  z^I\, \Theta^{-1}_{IJ}\, w^J} \, g \big) (x +z,y+w) \ . \nonumber
\eea
Introducing the shift operator $(\Tcal_a f)(x) := f(x +a)$ for any
$2d$-vector $a$, the twist element acting on $C^\infty(\Mcal)$ thus becomes
\be
\mathcal F
= \Big(\,\frac 1{\pi\,\hbar}\,\Big)^{2d}\,
\int_{\Mcal}\, \dd^{2d} z \ \int_{\Mcal}\,
\dd^{2d} w \ \big(\e^{\frac{\ii}{\hbar}\,
  z^I\, \Theta^{-1}_{IJ}\, w^J} \, \Tcal_{z} \big) \ot
\big( \e^{\frac{\ii}{\hbar}\,
  z^I\, \Theta^{-1}_{IJ}\, w^J} \, \Tcal_{ w}\big) \ .
\ee


\newsection{Nonassociative exterior differential calculus on $R$-space\label{calculus}}

The approach of Section~\ref{deformation} has the great virtue of
enabling a systematic development of exterior differential calculus on these
deformations. In this section we develop the basic ingredients
necessary for both an investigation of nonassociative quantum
mechanics within a phase space quantization formalism, which we
undertake in Section~\ref{NAQM}, as well as a putative formulation of field theories on the
nonassociative parabolic $R$-flux backgrounds.


\subsection{Covariant differential calculus\label{ndc}}

In this section we will use the cochain twist \eqref{mcF} to deform
the exterior algebra of differential forms $(\Omega^\bullet, \mu_\wg,
\dd)$, where $\Omega^\bullet:=\bigoplus_{n\geq 0}\, \Omega^n$ with
$\Omega^n=\Omega^n ({\cal M})$ the vector space of complex smooth
$n$-forms on ${\cal M}$, $\mu_\wg : \Omega^n \ot \Omega^m  \to
\Omega^{n+m}$ the usual exterior product
$\mu_\wedge(\omega\otimes\omega'\, ):= \omega\wedge\omega'$ and $\dd: \Omega^n \to
\Omega^{n+1}$ the exterior derivative with $\dd^2=0$.

We demand that the action
of the Hopf algebra $H=U(\mfg)$ from Section~\ref{lta} on
$(\Omega^\bullet, \mu_\wg, \dd)$ is covariant in the sense that
(c.f.~\eqref{tri})
\be
h\tri (\omega \wg \omega'\,)=\big(h_{(1)}\tri \omega\big) \wg \big(
h_{(2)}\tri \omega'\, \big)
\ee
and that $\dd$ is equivariant under the action of $H$ in the sense
that
\be
h \tri ( \dd\omega)=\dd( h\tri\omega) \ , \label{dequivariance}
\ee
for all $\omega, \omega' \in \Omega^\bullet$ and $h\in H$; in
  the framework of Section~\ref{categories}, these two
  conditions respectively mean that the structure maps
  $\mu_\wedge$ and $\dd$ are both morphisms in the category ${}_H\CCM$ of left $H$-modules. The action of $H$ on $\Omega^\bullet$ can be determined by finding the action on $\Omega^1$ and extending it to $\Omega^\bullet$ as an algebra homomorphism using the Leibniz rule
\be
\dd ( \omega \wg \omega' \, ) = \dd \omega \wg \omega' + (-1)^{\deg\omega} \, \omega \wg \dd \omega' \label{Leibniz1}
\ee
for all $\omega,\omega' \in \Omega^\bullet$. Employing
\eqref{dequivariance}, the action of $H$ on $\Omega^0$ given by
\eqref{rep} and \eqref{repM}, and the fact that $\dd$ commutes with
the Lie derivative along any vector field, we conclude that the action
of $H$ on $\Omega^\bullet$ is given by the Lie derivative $\mc L_h$
along elements $h\in H$. As previously the action is defined on
primitive elements of $H$ as an algebra homomorphism, i.e. $\mc
L_{\xi\, \xi'}:=\mc L_\xi\circ \mc L_{\xi'}$ for $\xi,\xi' \in \mfg
\subset U(\mfg)$, and it extends to a left action via linearity of the
Lie derivative and the Leibniz rule to get $\mc L_{h\, h'}=\mc
L_h\circ \mc L_{h'}$ for all $h,h' \in H$. Calculating this action on the generating $1$-forms gives
\be
M_{ij} \tri\dd x^k := \mc L_{M_{ij}}\big(\dd x^k \big) = \delta_j{}^k \,\dd p_i -\delta_i{}^k \, \dd p_j \ ,
\ee
with all other generators $\dd x^I$ invariant under the action of $H$. 

Following the methods of Section~\ref{td} (with $A=\Omega^\bullet$ and
$\mu_A=\mu_\wg$), we ensure that $\Omega^\bullet$ is covariant under
the action of $\mcH = (H_\mcF,\phi)$ by introducing a deformed
exterior product $\wg_\star$ on $\Omega^n [[\hbar]]\ot \Omega^m [[\hbar]] \to \Omega^{n+m}[[\hbar]]$ given by the formula 
\be 
\omega \wg_\star \omega' = \mu_\wg \big(\mcF^{-1} \tri (\omega \ot
\omega'\, ) \big) = \big(\mcF^{-1}_{(1)} \tri \omega\big)\wedge
\big(\mcF^{-1}_{(2)} \tri \omega'\, \big)
\ , \label{ws}
\ee
for all $\omega, \omega' \in \Omega^\bullet$. The exterior derivative
is still a derivation for the deformed exterior product and thus we
call the twisted exterior algebra $(\Omega^\bullet [[\hbar]],
\wg_\star , \dd)$ the \emph{nonassociative exterior differential calculus}. Using \eqref{ws} on the generating $1$-forms we find the relations
\be
\dd x^I \wg_\star \dd x^J = - \dd x^J \wg_\star \dd x^I = \dd x^I \wg \dd x^J \ ,
\ee
where $I,J \in \{1,\dots,2d\}$. We can again write \eqref{product of 3} in a more enlightening form for the case at hand as
\be
\begin{split}
( \omega \wg_\star \omega' \, ) \wg_\star \omega'' =\omega \wg_\star (
\omega' \wg_\star \omega'' \, ) + \sum_{n=1}^\infty \, \frac{1}{n!}\,
\Big(\, \frac{\hbar^2}{2}\, \Big)^n \, R^{i_1 j_1 k_1}\cdots R^{i_n j_n k_n}
\, \mc L_{i_1}\cdots \mc L_{i_n} (\omega) \ \wg_\star  \\
\wg_\star \ \big(\mc L_{j_1}\cdots \mc L_{j_n} (\omega'\, ) \wg_\star
\mc L_{k_1}\cdots \mc L_{k_n} (\omega''\, ) \big) \label{3omegas}
\end{split}
\ee
for all $\omega, \omega',\omega'' \in \Omega^\bullet$, where we have
abbreviated $\mc L_i :=\mc L_{\p_i}$. When  $\omega, \omega',\omega''
\in \Omega^1$, this formula takes an even simpler form given by
\be
\begin{split}
( \omega \wg_\star \omega' \, ) \wg_\star \omega'' =\omega \wg_\star (
\omega' \wg_\star \omega'' \, ) + \sum_{n=1}^\infty \, \frac{1}{n!}\, 
\Big(\, \frac{\hbar^2}{2}\, \Big)^n \, R^{i_1 j_1 k_1}\cdots R^{i_n
  j_n k_n} \,
 (\p_{i_1}\cdots \p_{i_n} \omega_L) \, \dd x^L \wg_\star \\  
 \wg_\star \big( (\p_{j_1}\cdots \p_{j_n} \omega'_M ) \,\dd x^M \wg_\star ( \p_{k_1}\cdots \p_{k_n} \omega''_N )\, \dd x^N \big) \ , \
\end{split}
\ee
where $i_l,j_l,k_l \in\{1,\dots,d\}$ and $L,M,N \in\{1,\dots,2d\}$. It follows that
\be
\big( \dd x^I \wg_\star \dd x^J\big)\wg_\star \dd x^K =  \dd x^I \wg_\star \big(\dd x^J\wg_\star \dd x^K \big) =: \dd x^I \wg_\star \dd x^J\wg_\star \dd x^K  \ ,
\ee
where $I,J,K \in \{1,\dots,2d\}$. 

The exterior product provides an $A$-bimodule structure on
$\Omega^\bullet$, where $A=(C^\infty({\cal M}), \mu_A)$, with right
and left action given by the pointwise multiplication of an $n$-form
by a function. Let us denote this action by $\rbtri$ and $\lbtri$
respectively; then covariance of the bimodule under the action of the
Hopf algebra $H$ means
\be
h\tri ( f \rbtri\omega)= \big(h_{(1)}\tri f\big)\rbtri \big(h_{(2)}\tri \omega \big) \ , \qquad  h\tri ( \omega\lbtri f )= \big(h_{(1)}\tri \omega \big)\lbtri\big(h_{(2)}\tri f \big) \ ,
\ee
for all $h\in H, \, f\in C^\infty({\cal M})$ and
$\omega\in\Omega^\bullet$. To ensure that the bimodule is covariant
under the action of the quasi-Hopf algebra $\mcH$ we must replace its action by the deformed right and left actions given respectively by the formulas
\be
f \rbtri_\star \omega = \lambda_\rbtri \big( \mcF^{-1} \tri (f \ot
\omega) \big) \ , \qquad \omega\, {}_\star{\lbtri} \, f = \lambda_\lbtri \big( \mcF^{-1} \tri (\omega \ot f) \big) \ , 
\ee
where $\lambda_\rbtri(f \ot \omega)=f\rbtri\omega=f\, \omega$ for all
$\omega \in \Omega^\bullet$ and $f \in C^\infty({\cal M})$, and
similarly for $\lambda_\lbtri : \Omega^\bullet \ot A \to
\Omega^\bullet$, which yields the deformed  $A_\star$-bimodule,
where $A_\star=(C^\infty({\cal M})[[\hbar]], \star)$. Since $f\rbtri
\omega=\omega\lbtri f $, here and throughout we will abuse notation
for the sake of simplicity by denoting $\rbtri_\star$ and ${}_\star{\lbtri}$ by $\star$ where no confusion arises. A short calculation then reveals the non-trivial bimodule relations between coordinates and $1$-forms given by
\be 
x^i \star \dd x^j =\dd x^j \star x^i + \mbox{$\frac{\ii \hbar}{2}$}
\,R^{ijk}\,\dd p_k \ ,
\ee
while all other left and right $A_\star$-actions coincide.


\subsection{Integration \label{integration}}

To compute quantum mechanical averages, and also to set up a
Lagrangian formalism for field theory, we need a suitable definition of integration $\int$ on $(\mc S({\cal M})[[\hbar]], \star)$, where $\mc S({\cal M})\subset C^\infty({\cal M})$ is the subalgebra of Schwartz functions on ${\cal M}=T^*M$.

Let us first notice that the star product \eqref{sp} satisfies 
\be 
f \star g = f \, g + \mbox{total derivative} \ . \label{total derivative}
\ee
This can be easily verified if we write the star product in the form \eqref{sstar}, and keep in mind that a total derivative in phase space includes both position and momentum derivatives. The order $\hbar^n$ term can then be written as
\be
\Theta^{I_1 J_1} \cdots \Theta^{I_n J_n} \,\big( \p_{I_1} \cdots
\p_{I_n} f \big) \, \big(\p_{J_1} \cdots \p_{J_n} g \big)= \p_{I_1} \cdots \p_{I_n} \big( \Theta^{I_1 J_1} \cdots \Theta^{I_n J_n} \, f \, \p_{J_1} \cdots \p_{J_n} g \big) 
\ee
since no momentum derivatives act on the upper left block of $\Theta $,
which means that \eqref{total derivative} is satisfied to all orders
in $\hbar$. Then the usual integration on ${\cal M}$ satisfies the \emph{2-cyclicity condition} 
\be 
\int_{{\cal M}}\, {\dd^{2d}x \ f \star g} =\int_{{\cal M}}\,
{\dd^{2d}x \ g \star f} = \int_{{\cal M}}\, {\dd^{2d}x \  f\, g} \label{2cyclicity}
\ee
for all $f,g \in \mc S ({\cal M})$, i.e. the standard integral on
$(\mc S({\cal M})[[\hbar]], \star)$ is 2-cyclic. 

In addition to the 2-cyclicity condition, the standard integral on $(\mc S({\cal M})[[\hbar]], \star)$ satisfies a cyclicity condition on the associator derived from the property 
\be
f \star (g \star h) = (f \star g) \star h +\trm{total derivative} \ , 
\ee
which easily follows from \eqref{Rassociator} and \eqref{Leibniz}. Hence the standard integral
also satisfies the property
\be
\int_{{\cal M}}\, {\dd^{2d}x \ f \star (g \star h)} = \int_{{\cal
    M}}\, {\dd^{2d}x \ (f \star g) \star h} =:\int_{{\cal
    M}}\, {\dd^{2d}x \ f \star g \star h} 
\label{3cyclicity}
\ee
for all $f,g,h \in \mc S({\cal M})$, which we call the
\emph{3-cyclicity condition}; this property was also
derived in~\cite{Mylonas:2012pg} from a slightly different perspective.

Our proofs of 2-cyclicity (\ref{2cyclicity}) and 3-cyclicity (\ref{3cyclicity}) are only carried out for constant $R$-flux in the standard coordinate frame. As changing coordinates
alters this situation, let us comment briefly on what restrictions are generically required. As we discuss at greater length in Section~\ref{NPstructures}, the antisymmetric tensor $R^{ijk}$ defines a 3-bracket which is Nambu-Poisson, and this property is invariant under change of coordinates. A Nambu-Poisson
tensor implies locally a foliation of the underlying manifold, while our proofs require this
property globally. It follows that 2-cyclicity and 3-cyclicity hold whenever $R$ is
a Nambu-Poisson trivector that defines a global foliation of $M$.

The 2-cyclicity condition does \emph{not} generally guarantee the usual
cyclicity property involving integration of $n$-fold star products of
functions. This is because the star product is nonassociative and thus
a bracketing for the star product of $n$ functions has to be
specified. Once this is done one cannot freely move functions
cyclically under integration using \eqref{2cyclicity} as one would
normally do in the associative case; instead the 3-cyclicity condition
\eqref{3cyclicity} can be used to rebracket the integrated expression
and to investigate its equivalence with expressions involving
different bracketings. In general, the total number of ways to bracket
a star product of $n$ functions is given by the Catalan number $C_{n-1}$, where
\be
C_n = \frac{1}{n+1} \, {2n \choose n} = \frac{(2n)!}{n! \, (n+1)!} \label{Cn}
\ee
for $n\ge 0$. Starting from the integral
\be
\int_{{\cal M}}\, {\dd^{2d}x \ f_1 \star \big( f_2 \star (f_3 \star
  (\cdots \star f_n) \cdots )\big)}  \ , \label{nfold}
\ee
one can prove that it is equal to a number of different bracketings
but always inequivalent to any other bracketing of the form $f_1 \star
(\trm{bracketed expression})$, i.e. under 3-cyclicity the distinct
ways of bracketing an integrated $n$-fold star product of functions are organised into $C_{n-2}$ classes, one for each different bracketing where $f_1$ is free at the front. For example, for $n=4$ there are five different bracketings out of which two are of the form \eqref{nfold} and thus we have two different classes of equivalent bracketings, namely
\be
\int_{{\cal M}}\, {\dd^{2d}x \ f_1 \star \big( f_2 \star (f_3 \star
  f_4) \big)} = \int_{{\cal M}}\, {\dd^{2d}x \ (f_1 \star f_2) \star
  (f_3 \star f_4)} = \int_{{\cal M}}\, {\dd^{2d}x \ \big(( f_1 \star f_2) \star f_3 \big)\star f_4} 
\ee 
and
\be
\int_{{\cal M}}\, {\dd^{2d}x \ f_1 \star \big( (f_2 \star f_3) \star
  f_4 \big)} = \int_{{\cal M}}\, {\dd^{2d}x \ \big( f_1 \star (f_2 \star f_3) \big) \star f_4} \ .
\ee

A graded 2-cyclicity condition could also be derived for the deformed
exterior product, provided that we can write an equation similar to
\eqref{total derivative} for it. This seems
complicated since generally \eqref{ws} cannot be written explicitly in
closed form, but fortunately there is a way around this problem: We
can use the result of~\cite{Aschieri:2009ky} where it was shown that
if the identity
\be
U_\mcF = \mcF_{(1)} \, S\big(\mcF_{(2)}\big) = 1_H \label{SFF}
\ee
holds, where $U_\mcF=\mu\circ(\Id_H\ot S)(\mcF )$ and $S$ is the
antipode, then standard integration on $(\Omega^\bullet[[\hbar]] ,
\wg_\star, \dd)$ is graded 2-cyclic. This is always true for
abelian twists but does not hold in general; however, in our case the
twisted antipode $S_\mcF$ coincides with $S$. It is then
straightforward to demonstrate \eqref{SFF} on $\mc S({\cal M})$ by
using the representation of primitive elements of $H$ on functions in
\eqref{mcF} and antisymmetry of $R^{ijk}$. Hence we conclude that for
the nonassociative exterior differential calculus
$(\Omega^\bullet[[\hbar]] , \wg_\star, \dd)$ the graded 2-cyclicity condition 
\be
\int_{{\cal M}}\, {\omega \wg_\star \omega'}
=(-1)^{\trm{deg}(\omega)\, \trm{deg}(\omega'\, )}\, \int_{{\cal M}}\,
{\omega' \wg_\star \omega} = \int_{{\cal M}}\, {\omega \wg \omega'}  \label{gradedcyclicity}
\ee
is indeed satisfied. 

The 3-cyclicity condition \eqref{3cyclicity} can also be generalized by noticing that similarly to $\p_i$ being a derivation for the nonassociative star product by \eqref{Leibniz}, the Lie derivative $\mc L_i$ is a derivation of the deformed exterior product $\wg_\star$ by \eqref{Delta} and the discussion that followed \eqref{Leibniz}, and since $[\mc L_i, \mc L_j] = \mc L_{[\p_i,\p_j]}=0$ by \eqref{3omegas} one has
\be
\omega \wg_\star
(\omega' \wg_\star \omega''\, ) = (\omega \wg_\star \omega'\, ) \wg_\star \omega'' +\trm{total Lie derivative} \ . 
\ee
Since $\int_{{\cal M}}\, {\mc L_i(\omega)}=0$ for all $\omega \in \Omega^\bullet$ we thus get the generic 3-cyclicity condition 
\be
\int_{{\cal M}}\, {\omega \wg_\star (\omega' \wg_\star \omega''\, )} =
\int_{{\cal M}}\, {(\omega \wg_\star \omega'\, ) \wg_\star \omega''}=:
\int_{{\cal M}}\, {\omega \wg_\star \omega' \wg_\star \omega'' } \label{omegatriplecyclicity}
\ee
for all $\omega, \omega', \omega'' \in \Omega^\bullet$, generalizing \eqref{3cyclicity} which is the $\Omega^0$ case.


\newsection{Nonassociative quantum mechanics\label{NAQM}}

The standard formulation of quantum mechanics is 
based on linear operators acting on a separable Hilbert space and the corresponding operator algebras are by construction associative. Nevertheless, it turns out that the mathematical tools and structures that we have developed in this paper do in fact allow for a direct quantitative discussion of nonassociativity in quantum mechanics, adding to the more qualitative arguments that can already be found in the literature.
The lack of associativity alters the theory of quantum mechanics
drastically, but against all odds a consistent formulation is
apparently indeed possible. To the best of our knowledge the bulk of
the material presented in this section is new.


\subsection{Nambu-Heisenberg brackets, star products and compositions}

The task of formulating a nonassociative version of quantum mechanics is closely related to the quest of quantizing Nambu-Poisson brackets (see e.g.~\cite{DeBellis:2010pf, Saemann:2012ab} and references therein). A natural choice would be the Jacobiator of operators, but it obviously vanishes for associative operator algebras. As a work-around, a Nambu-Heisenberg bracket was introduced by Nambu as half of a Jacobiator~\cite{Nambu:1973qe}
\be
[A,B,C]_\text{NH} = A \, B \, C + C \, A \, B + B \, C \, A - B \, A
\, C   - A \, C \, B - C \, B \, A \ .
\ee
It is straightforward to evaluate the Nambu-Heisenberg bracket on
coordinate functions for any of our star
products.\footnote{Nambu-Heisenberg brackets have been previously
  investigated in the context of phase space quantum mechanics based
  on \emph{associative} star products in~\cite{Zachos:2003md}.} For
instance, for the associative constant $\bar p$ star product \eqref{ts} we find
\be \label{NHts}
[x^i,x^j,x^k]_\text{$\ts$-NH} =\ii \hbar \, \big(R^{ijl} \, \bar p_l \, x^k   + R^{jkl} \, \bar p_l \, x^i  +  R^{kil} \, \bar p_l \, x^j\big) \ .
\ee
Nambu suggested to consider nonassociative algebras for the quantization of his bracket. We do have the tools now to study this proposal.
In the nonassociative case, we need to specify which operators are
multiplied first. We choose by default the first pair and write
\be
[A,B,C]_\text{NH} = [A,B] \, C + [C,A] \, B + [B,C] \, A \ ,
\ee
where $[A,B]\, C := (A\, B)\, C - (B\, A)\, C$.
For the nonassociative star product \eqref{sp}, evaluated on a triple of coordinate functions, this gives
\be \label{NHstar}
[x^i,x^j,x^k]_\text{$\star$-NH} = \ii \hbar \, \big( R^{ijl} \, p_l \star x^k   + R^{jkl} \, p_l \star x^i  +  R^{kil} \, p_l \star x^j \big) \ .
\ee
The opposite Nambu-Heisenberg bracket
\be
[A,B,C]'_\text{NH} = C \, [B,A]  + B \, [A,C]  + A \, [C,B]
\ee
is in general no longer equal to minus the original Nambu-Heisenberg
bracket. Their sum gives the Jacobiator
\be
[[A,B,C]] := [[A,B], C] + [[C,A], B] + [[B,C], A]  = [A,B,C]_\text{NH} + [A,B,C]'_\text{NH} \ .
\ee
For the nonassociative star product \eqref{sp}, evaluated on a triple
of coordinate functions, we obtain the non-zero Jacobiator (c.f.~\eqref{jacobiator})
\eq \label{Jacobia}
[[x^i, x^j, x^k]]_\star = \ii \hbar \, \big(R^{ijl} \, [p_l, x^k]_\star   + R^{jkl} \, [p_l, x^i]_\star  +  R^{kil} \, [p_l, x^j]_\star \big)
= 3 \, \hbar^2 \, R^{ijk} \ 
\ee
as a more convincing candidate than \eqref{NHts} or \eqref{NHstar} for a quantized Nambu-Poisson bracket.

An indirect approach to nonassociative quantum mechanics can be based
on the family of associative star products $\ts$ for constant
$\tp$-slices and the mappings that link them, and to $\star$ by
twists, very much in the spirit of describing a general manifold in terms of Euclidean spaces by local coordinate charts and transition functions. A regular operator/Hilbert space approach to nonassociative quantum mechanics can in fact be based on standard canonical quantization and the twist~\eqref{FR} from the Moyal-Weyl product \eqref{moyal} to the nonassociative product \eqref{sp}; after quantization, the twist is expressed in terms of operators acting on a suitable Hilbert space.

Instead of these solid but indirect approaches to nonassociative quantum mechanics, we shall pursue a more direct approach:
The phase space formulation of  quantum mechanics \cite{Moyal:1949sk} is powerful enough to study nonassociative quantum mechanics \emph{in situ} (see e.g.
\cite{Zachos:2001ux} and references therein).  Observables are
implemented as real functions on phase space, states are represented
by pseudo-probability Wigner-type density functions, and
noncommutativity of operators enters via a star product of functions, which is the deformation quantization of a classical Poisson structure.

Let us start by introducing some convenient notation and conventions.
We introduce the \emph{compositions} $\circ$ and $\cirp$ by
\be
(A \circ B)\star C := A \star (B \star C) \ ,  \qquad  C\star (A \cirp B) := (C \star A) \star B
\ee
for all $A, B, C \in C^\infty({\cal M})[[\hbar]]$. The compositions are related by complex conjugation: $(A \circ B)^* = B^* \cirp A^*$ and \emph{vice versa}.
We choose the convention that $\cirp$ is evaluated before $\circ$ in
all expressions that involve both compositions, so that\footnote{This
  convention looks asymmetric, but as long as we are just computing
  expectation values, it gives the same results as the alternative
  convention as a consequence of 3-cyclicity. Physically this is a remnant of operator-state duality. In the context of  time-evolution and similar transformations this duality is, however, no longer a symmetry in the nonassociative setting.}
\be
(A \circ B) \star (C \cirp D) := \big((A \circ B)\star C \big) \star D = \big( A \star (B \star C)\big) \star D \ .
\ee
The compositions can be extended to an arbitrary number of functions and are by construction associative.
Like the star products that we are considering in this paper, the compositions are noncommutative and unital: $1 \circ A = A = A \circ 1$.
For an associative algebra, $\circ$ would just be the product in that algebra. However, in the nonassociative case $A \circ B$ cannot even generally be replaced by some suitable element of the algebra $(C^\infty({\cal M})[[\hbar]],\star)$, because if this were possible then $(A \circ B) \star 1 = A \star (B \star 1) = A \star B$ would imply $A \circ B$ = $A \star B$ and thus
$(A \star B) \star C = (A \circ B) \star C = A \star (B \star C)$ for
all $C \in C^\infty({\cal M})[[\hbar]]$. In a nonassociative algebra
this is obviously not true for all $A, B \in C^\infty({\cal M})[[\hbar]]$. There are, however, some notable exceptions, e.g. $x^i \circ x^i = x^i \star x^i = (x^i)^2$ and
$p_i \circ p_i = p_i \star p_i = (p_i)^2$.


\subsection{States, operators and eigenvalues}
\label{sec:states}

States map observables to numbers, which are interpreted as expectation values and link theory to experiment. For this purpose one requires convexity, reality, unit trace, and positivity properties. The latter property is particularly difficult to implement in a nonassociative setting. A definition that ultimately fulfills all these requirements is as follows.
A \emph{state} $\rho$ is an expression of the form
\be \label{rho}
\rho = \sum_{\alpha=1}^n \, \lambda_\alpha \,\psi_\alpha \cirp \psi_\alpha^* \ ,
\ee
where $n \geq 1$, $\lambda_\alpha > 0$, $\sum_{\alpha=1}^{n}\,
\lambda_\alpha = 1$, and $\psi_\alpha$ are complex-valued \emph{phase
  space wave functions}, which are normalized as
\be
\intps |\psi_\alpha|^2 = 1 \ ,
\ee
but are not necessarily orthogonal.\footnote{Using the familiar
  language of quantum mechanics, we refer to complex-valued functions
  on phase space that are multiplied by star products as ``operators''
  and to real-valued functions on phase space that are associated to
  something that can in principle be measured as ``observables''. The
  phase space wave functions $\psi_\alpha$ and their complex
  conjugates $\psi_\alpha^*$ should not be confused with state vector
  kets or bras. The corresponding objects in ordinary quantum
  mechanics are normalized but otherwise arbitrary operators that are
  not necessarily related to rank one projectors.} For any two states
$\rho_1$ and $\rho_2$, the convex linear combination $\rho_3 =
\lambda\, \rho_1 + (1-\lambda) \, \rho_2$ with $\lambda \in [0,1]$ is again a state. The space of states is thus a convex set, whose extrema we call pure states. A necessary (but not sufficient) condition for a state to be pure is that it is of the form $\rho = \psi \cirp \psi^*$.
Given a state $\rho$, the expectation value of a function on phase space (``operator'') $A$  is obtained by the phase space integral
\be \label{expvalue}
\begin{split}
\langle A \rangle :=& \, \intps A \star \rho \\[4pt]
=& \, \sum_{\alpha=1}^n\, \lambda_\alpha \ \intps A \star (\psi_\alpha
\cirp \psi_\alpha^*) \\[4pt] 
=& \, \sum_{\alpha=1}^n \, \lambda_\alpha \ \intps (A \star \psi_\alpha)
\star \psi_\alpha^* \\[4pt]
 =& \, \sum_{\alpha=1}^n \, \lambda_\alpha \ \intps \psi_\alpha^* \star (A \star \psi_\alpha) 
= \sum_{\alpha=1}^n\, \lambda_\alpha \ \intps \psi_\alpha^* \, (A \star \psi_\alpha) \ ,
\end{split}
\ee
where we have used 2-cyclicity.
Using 3-cyclicity and the fact that complex conjugation acts anti-involutively on the star product,  $(A \star \psi)^* = \psi^* \star A^*$, we find
\be
\langle A \rangle^* = \sum_{\alpha=1}^n \, \lambda_\alpha \ \intps (A \star \psi_\alpha)^* \star \psi_\alpha
= \sum_{\alpha=1}^n \, \lambda_\alpha \ \intps \psi_\alpha^* \star (A^* \star \psi_\alpha) = \langle A^* \rangle \ .
\ee
Observables (i.e. real-valued functions on phase space $A^* = A$) therefore  have real expectation values as desired. We will later show that 3-cyclicity also ensures reality of eigenvalues. Thanks to 3-cyclicity, our approach to nonassociative quantum mechanics is thus not affected by a previously proposed no-go theorem~\cite{Dzhunushaliev:2007wu}.

Expectation values can also be computed for star products of functions
(because star products of functions are again functions). The
definition of expectation value can be further extended to
compositions of operators as
\be
\begin{split}
\langle A_1 \circ A_2 \circ \cdots \circ A_k \rangle \,
&=  \sum_{\alpha=1}^n \, \lambda_\alpha \ \intps  (A_1 \circ A_2 \circ
\cdots \circ A_k) \star \psi_\alpha \cirp \psi^*_\alpha \\[4pt] 
&= \sum_{\alpha=1}^n\, \lambda_\alpha \ \intps \big[A_1 \star \big(A_2
\star \big(\cdots  \star (A_k \star \psi_\alpha)\cdots\big) \big) \big] \star \psi^*_\alpha \ .
\end{split}
\ee
Positivity is a tricky concept in the nonassociative setting. In terms
of our definition of a state, it is realized for any state $\rho$ and
any function on phase space $A$ as
\eq
\begin{split}
\langle A^* \circ A \rangle \,
&= \sum_{\alpha=1}^n \, \lambda_\alpha \ \intps \psi^*_\alpha \star \big(A^* \star (A \star \psi_\alpha)\big) \\[4pt]
&= \sum_{\alpha=1}^n \, \lambda_\alpha \ \intps (\psi^*_\alpha \star A^*) \star (A \star \psi_\alpha) \\[4pt]
&= \sum_{\alpha=1}^n \, \lambda_\alpha \ \intps (A \star \psi_\alpha)^* \, (A \star \psi_\alpha) \\[4pt]
&= \sum_{\alpha=1}^n \, \lambda_\alpha \ \intps |A \star \psi_\alpha|^2 \geq 0 \ ,
\end{split}
\ee
where we have used 2-cyclicity, 3-cyclicity and anti-involutivity with respect to complex conjugation.  With a similar computation we see that
\be
(A,B) := \langle A^* \circ B\rangle
= \sum_{\alpha=1}^n \, \lambda_\alpha \ \intps (A \star \psi_\alpha)^* \, (B \star \psi_\alpha)
\ee
defines a semi-definite sesquilinear form for any given state $\rho$. This will be the basis of the derivation of uncertainty relations below, because it implies the Cauchy-Schwarz inequality
\be \label{Cauchy-Schwarz}
\big|(A,B)\big|^2 \leq  (A,A)\, (B,B) \ .
\ee
Using 3-cyclicity, the expectation value \eqref{expvalue} of a single operator (no compositions) can be rewritten in terms of
a \emph{state function} $S_\rho = \sum_{\alpha=1}^n \, \lambda_\alpha
\, \psi_\alpha \star \psi_\alpha^*$ as
\be
\langle A \rangle =
 \sum_{\alpha=1}^n \, \lambda_\alpha \ \intps (A \star \psi_\alpha)\star  \psi_\alpha^*
= \sum_{\alpha=1}^n \, \lambda_\alpha \ \intps  A \star (\psi_\alpha \star  \psi_\alpha^*)
= \intps A \, S_\rho \ .
\ee
The state function $S_\rho$ is a real function on phase space that is
normalized as
\be
\langle 1 \rangle = \intps S_\rho = \sum_{\alpha=1}^n \,
\lambda_\alpha \ \intps |\psi_\alpha|^2 = 1 \ ,
\ee
but is not necessarily non-negative everywhere. It plays the role of a quasi-probability distribution function, like the Wigner function in the associative case.
However, unlike the associative case, we cannot formulate the theory entirely in terms of the state function $S_\rho$, but rather we also need to frequently refer to the phase space wave functions $\psi_\alpha$.

A function (``operator'') $A$ can have eigenfunctions $f$ (with respect to $\star$-multiplication) with eigenvalues $\lambda \in \mathbb C$:
$A \star f = \lambda \, f$. Complex conjugation implies  $f^* \star
A^* = \lambda^* \, f^*$. We can show that real functions $A=A^*$ have
real eigenvalues, but this fact is not quite as straightforward as in
the associative case. We have
\be
f^* \star (A \star f) - (f^* \star A) \star f = (\lambda -
\lambda^*)\, (f^* \star f) \ .
\ee
The left-hand side of this equation is non-zero in general, but it vanishes after integrating over phase space and using 3-cyclicity. We obtain
\be
(\lambda-\lambda^*) \, \intps f^* \star f = (\lambda-\lambda^*) \, \intps |f|^2 = 0 \ .
\ee
The integral is non-zero unless $f$ is identically equal to zero and
therefore $\lambda = \lambda^*$ as desired. Using similar
manipulations, we can show that eigenfunctions are orthogonal if they
correspond to distinct eigenvalues. In the nonassociative case we need
to distinguish eigen-state functions and eigen-wave functions (unless
we integrate and use 3-cyclicity): $A \star \psi = \lambda \, \psi$
does not necessarily imply
$A \star S_\rho = \lambda \, S_\rho$, where $\rho = \psi \cirp \psi^*$ and  $S_\rho= \psi \star \psi^*$, because
$(A \star \psi) \star \psi^* \neq A \star (\psi \star \psi^*)$ in general.

All definitions here and in the following are consistent with the
associative limit of phase space quantum mechanics. The nonassociative
case is more restrictive and in a way it teaches us also something
about ordinary phase space quantum mechanics. We have attempted to
keep all definitions as general as possible. Depending on the intended
application, further restrictions may be necessary; for example, it is
natural to require states to be symmetric: $\rho = \rho'$, where
$\rho$ is as given in \eqref{rho} and $\rho' = \sum_{\alpha=1}^n \, \lambda_\alpha \,\psi_\alpha^* \cirp \psi_\alpha$.


\subsection{Uncertainty relations, area and volume operators}

A pair of operators that do not commute cannot have a complete set of common eigenstates; a pair of operators with a central non-zero commutator do not have any simultaneous eigenstates. These  well-known facts of quantum mechanics are important for measurements and can also be verified for nonassociative phase space quantum mechanics. A new feature is that analogous statements hold for any triple of operators that do not associate. Let us illustrate this for phase space coordinate functions
$x^I \in \{x^1,\ldots,x^d,p_1,\ldots,p_d\}$ with commutator and associator
\be
x^I \star x^J - x^J \star x^I = \ii\hbar \, \Theta^{IJ} \ , \qquad
(x^I \star x^J) \star x^K - x^I \star (x^J  \star x^K) = \mbox{$\frac{\hbar^2}2$}\, R^{IJK} \ ,
\ee
respectively, where $R^{IJK} := \partial_K \Theta^{IJ}$ is constant and non-zero (and then equal to $R^{ijk}$) only for (selected) configuration space coordinates, c.f. \eqref{Theta} and \eqref{Rassociator}.
Let us assume that a pair of phase space coordinates $x^I$ and $x^J$
with $I\neq J$ have a common (normalized) eigen-state function $S$: $x^I
\star S = \lambda^I \, S$ and $x^J \star S = \lambda^J \, S$. Using 3-cyclicity, this implies
\be
\big\langle [x^I,x^J]_\star \big\rangle = 
\intps \big(x^I \star (x^J \star S) - x^J \star (x^I \star S)\big)
= \lambda^I \,\lambda^J - \lambda^J \,\lambda^I  = 0 \ ,
\ee
and hence $x^I$ and $x^J$ with $\big\langle \Theta^{IJ} \big \rangle
\neq 0$ cannot have a common eigen-state function $S$. Let us now
assume that a triple of phase space coordinates $x^I$, $x^J$, and $x^K$
have a common eigen-state function $S$ with eigenvalues $\lambda^I$,
$\lambda^J$, and $\lambda^K$.
Using 3-cyclicity repeatedly we find
\be
\begin{split}
\intps \big((x^I \star x^J)\star x^K\big) \star S \, &= \intps (x^I \star
x^J)\star (x^K \star S) \\[4pt]
&= \lambda^K \, \intps (x^I \star x^J)\star S \\[4pt]
&= \lambda^K \, \intps x^I \star (x^J\star  S) = \lambda^K \,
\lambda^J \, \lambda^I \ ,
\end{split}
\ee
while using 2-cyclicity, 3-cyclicity and the fact that $\lambda^I$ must be real we find similarly
\be
\intps \big(x^I \star (x^J\star x^K)\big) \star S  = \intps (x^J \star
x^K)\star (S \star x^I) = \lambda^I \, \lambda^K \, \lambda^J \ .
\ee
Taking the difference of the two expressions implies
\be
\mbox{$\frac{\hbar^2}2$} \, R^{IJK} = \lambda^K \, \lambda^J \,
\lambda^I - \lambda^I \, \lambda^K \, \lambda^J = 0 \
\ee
and we arrive at the striking result that coordinates $x^i$, $x^j$ and
$x^k$ which do not associate, i.e. for which $R^{ijk} \neq 0$, cannot
have a common eigen-state function $S$; whence they cannot be measured
simultaneously with arbitrary precision. This is a clear sign of a
coarse-graining (quantization) of space in the presence of $R$-flux.

Let us now turn to the study of uncertainties. In the definition of the uncertainties, we \emph{a priori} face the problem of having to make a choice
between using expectation values based on phase space wave functions (with the advantage of the availability of inequalities) and state functions (with computational advantages).
In the computation of uncertainties for phase space coordinates, this luckily does not play a role because $x^I \circ x^I = x^I \star x^I$
and thus
\be
0 \leq \sum_{\alpha=1}^n \, \lambda_\alpha \ \intps \psi^*_\alpha \star \big(x^I \star (x^I \star \psi_\alpha)\big) = \intps (x^I \star x^I) \star S_\rho
= \big\langle (x^I)^{\star 2}\big\rangle \ .
\ee
Without ambiguity, we can therefore define the uncertainty as usual in
terms of the expectation value of the square of the shifted coordinate
$\widetilde x^I := x^I - \langle x^I \rangle$ as
\be
\Delta x^I := \sqrt{\big\langle (\widetilde x^I)^{\star 2}\big\rangle} = \sqrt{\big\langle (x^I)^{\star 2}\big\rangle - \big\langle x^I\big\rangle^2  } \ .
\ee
This uncertainty is zero for eigen-state functions ($x^I \star S = \lambda\, S$) as well as for eigen-wave functions
($x^I \star \psi = \lambda \, \psi$). The function $x^I$ is real and the uncertainty can thus be rewritten as
\be
(\Delta x^I)^2 = \big\langle \widetilde x^I \star \widetilde x^I\big\rangle =  \big\langle \widetilde x^I \circ \widetilde x^I \big\rangle = \big(\widetilde x^I \, ,\widetilde x^I \big) \ .
\ee
Using the Cauchy-Schwarz inequality \eqref{Cauchy-Schwarz}, and decomposing into imaginary and real parts we get
\be
(\Delta x^I)^2 \, (\Delta x^J)^2 \geq  \big|(\widetilde x^I,\widetilde x^J)\big|^2 = \mbox{$\frac14$}\, \big|\big\langle [x^I,x^J]_\circ \big\rangle\big|^2 + \mbox{$\frac14$}\, \big|\big\langle \{\widetilde x^I,\widetilde x^J\}_\circ \big\rangle\big|^2 \ ,
\ee
where $[A,B]_\circ := A \circ B - B \circ A$ and $\{A,B\}_\circ := A \circ B + B \circ A$.
Ignoring the last term yields a Born-Jordan-Heisenberg-type uncertainty relation
\be
\Delta x^I \, \Delta x^J \geq \mbox{$\frac12$} \, \big|\big\langle [x^I,x^J]_\circ \big\rangle\big| \ .
\ee
To proceed from here, we need to distinguish several cases:  Whenever one of the phase space coordinates is a momentum $p_i$, nonassociativity does not play a role in the sense that then ``$\circ = \star$'', i.e.\
$[p_i,p_j]_\circ = [p_i,p_j]_\star = 0$ and
$[p_i,x^j]_\circ = [p_i,x^j]_\star = - \ii\hbar \,\delta_i{}^j$, and therefore
\be
\Delta p_i \, \Delta p_j \geq 0 \ , \qquad
\Delta x^i \, \Delta p_j \geq \mbox{$\frac\hbar 2$} \, \delta^i{}_j
\ee
as in ordinary quantum mechanics. The non-trivial uncertainty relation for a pair of coordinates $x^i$ and $x^j$ is new and requires a more complicated computation. With the help of the associator \eqref{Rassociator} we can express $[x^i,x^j]_\circ$ in terms of a star commutator and obtain the surprising result
\be
\begin{split}
[x^i,x^j]_\circ \star \psi  :=& \, x^i\star(x^j \star \psi) -
x^j\star(x^i \star \psi) \\[4pt] 
=& \, [x^i,x^j]_\star \star \psi - \hbar^2 \, R^{ijk} \, \partial_k \psi
\\[4pt] 
=&\, {\ii\hbar}\,R^{ijk}\,\big(p_k \star \psi  + \ii\hbar \, \partial_k
\psi \big)\\[4pt] 
=& \, {\ii\hbar}\,R^{ijk}\,\big(p_k \, \psi - \mbox{$\frac{\ii\hbar}2$} \, \partial_k \psi + \ii\hbar \, \partial_k \psi \big)
= {\ii\hbar}\,R^{ijk} \, \psi \star p_k \ ,
\end{split}
\ee
that is, the momentum operator ends up on the ``wrong'' side of $\psi$. Using this result, we obtain an uncertainty relation for position measurements
\be \label{newunc}
\Delta x^i \, \Delta x^j \geq \mbox{$\frac\hbar 2$}\, \big| R^{ijk} \,
\langle p_k \rangle'\big| \ ,
\ee
where $\langle p_k \rangle'$ is the expectation value of $p_k$
computed with respect to the \emph{opposite} state $\rho'$. Only for
symmetric or antisymmetric states, i.e. $\rho' = \pm \, \rho$, will
this be equal to the standard expectation value $\langle p_k\rangle$,
and one should consider adding this requirement to the definition of a
state. The new uncertainty relation \eqref{newunc} features uncertainties for position measurements in directions transverse to momentum, while the usual Heisenberg uncertainty relation relates uncertainties of position and momentum in the same direction.

It is tempting to interpret the left-hand side of \eqref{newunc} as an
area uncertainty that grows linearly with transverse momentum, but
this is misleading: The position uncertainty relation makes a
prediction for the average outcome of many identically prepared
experiments in which \emph{either} $x^i$ \emph{or} $x^j$ is
measured. In none of these (Gedanken-)experiments are positions in two
different directions  measured simultaneously (or one shortly after
the other), but this would be required for a genuine area
uncertainty. An analogous criticism applies  to the superficial
interpretation of Heisenberg uncertainty as an uncertainty of phase
space areas in ordinary quantum mechanics. To remedy the situation, we
shall define an area operator whose expectation value can be computed
and interpreted as fundamental area measurement uncertainty (or
minimal area). The approach generalizes to higher dimensional objects
and we will also derive a fundamental volume measurement uncertainty,
which results from the nonassociativity of coordinate
functions.\footnote{See also \cite{Bahns:2010um} for area and volume
  operators in an associative setting.}

The oriented area spanned by two segments $\delta\vec r_1$ and $\delta\vec r_2$ in three-dimensional Euclidean space is given by the vector product $\delta\vec r_1 \times \delta\vec r_2$, while the volume spanned by three segments $\delta\vec r_1$, $\delta\vec r_2$ and $\delta\vec r_3$ is the triple scalar product $\delta\vec r_1 \cdot(\delta\vec r_2 \times \delta\vec r_3)$. This can easily be generalized to higher dimensional parallelepipeds and to embedding spaces of arbitrary dimension. The most convenient description of these areas and (higher dimensional) volumes  for our purposes is in terms of antisymmetrized sums of products of components of displacement vectors $\delta \vec r$. For the sake of generality we might as well consider displacements in phase space. For the description of quantum uncertainties we replace all of the displacement vectors by the single displacement vector (of operators) $\vec {\widetilde x} = \vec x - \langle \vec x\, \rangle$, and promote commutative pointwise multiplication to the noncommutative and nonassociative star product~$\star$. Furthermore, we would like to construct observables, i.e. real functions on phase space. Taking all this into consideration, the appropriate area (uncertainty) operator in directions $x^I$, $x^J$ is
\be
A^{IJ} = \mathfrak{Im}\big([\widetilde x^I, \widetilde x^J]_\star\big) = -\ii\big({\widetilde x^I} \star {\widetilde x^J} - {\widetilde x^J} \star {\widetilde x^I}\big)
\ee
and the volume (uncertainty) operator in directions $x^I$, $x^J$,
$x^K$ is (c.f.~\eqref{NHstar} and \eqref{Jacobia})
\be
V^{IJK} = \mathfrak{Re} \big( [\widetilde x^I, \widetilde x^J, \widetilde x^K]_\text{$\star$-NH}\big) = 
\mbox{$\frac12$} \, \big[\big[\widetilde x^I, \widetilde x^J, \widetilde x^K\big]\big]_\star \ .
\ee
The expectation values of these (oriented) area and volume operators are easily computed to be
\be
\langle A^{IJ} \rangle = \hbar \, \big\langle \Theta^{IJ}\big \rangle \ ,
\qquad \langle V^{IJK} \rangle = \mbox{$\frac32$} \, \hbar^2 \, R^{IJK} \ , 
\ee
with three interesting special cases
\be
\langle A^{x^i,p_j} \rangle = \hbar \, \delta^i{}_j \ , \qquad \langle A^{ij} \rangle = \hbar \, R^{ijk} \,\langle p_k\rangle
\ , \qquad \langle V^{ijk} \rangle = \mbox{$\frac32$} \, \hbar^2\, R^{ijk} \ . \label{uncertainties}
\ee
The first expression describes phase space cells with area $\hbar$. The second expression illustrates an area uncertainty proportional to the magnitude of the transverse momentum.
The third expression indicates a minimal resolvable volume of order
$\frac32 \,\hbar^2\, |R|$ due to nonassociativity-induced position
measurement uncertainties (here $|R|$ is a generalized determinant of
the antisymmetric 3-tensor $R^{ijk}$). Uncertainties similar to
\eqref{uncertainties} have appeared previously
in~\cite{Lust:2010iy,Blumenhagen:2010hj}, and here we have provided a
concrete and rigorous derivation of them as expectation values of
area and volume operators.


\subsection{Dynamics and transformations}

Let us close this section with some remarks on dynamics, and similar state and operator transformations in nonassociative quantum mechanics. Time evolution and other transformations should leave the structure of the theory intact. In particular notions of positivity, normalization of probabilities and reality should be preserved. Observables (i.e. real functions on phase space) should be mapped to observables and (pure) states to (pure) states.
As in ordinary quantum mechanics, there are two approaches that fulfill all these requirements. In the nonassociative case the two approaches are, however, no longer equivalent.

A Schr\"odinger-type approach focuses on evolution equations for the phase space wave functions. The starting point is the phase space Schr\"odinger equation
\be \label{Schroedinger}
\ii \hbar \, \frac{\partial\psi }{\p t} = \mathcal H \star \psi \ ,
\ee
which applies to all $\psi_\alpha$ and $\psi_\alpha^*$ in the state
$\rho$ (c.f.~\eqref{rho}), and where the Hamiltonian $\mathcal H$ is a real function on phase space. Observationally, only the time evolution $\frac{\partial}{\p t}\langle A \rangle$ of expectation values is relevant. It can be computed either from the Schr\" odinger equation~\eqref{Schroedinger} or equivalently from the time evolution equation for operators and compositions of operators given by
\be \label{Heisenberg1}
\frac{\partial\alpha}{\p t}  = \frac \ii\hbar \, [\mathcal H,\alpha]_\circ \ ,
\ee
where $\alpha =$ $A$ or $A \circ B$ or $A \circ B \circ C$ etc. The $\circ$-commutator in \eqref{Heisenberg1} is a $\circ$-derivation,
\be
[\mathcal H,A \circ B]_\circ = [\mathcal H,A]_\circ \circ B + A \circ [\mathcal H,B]_\circ \ ,
\ee
and thus $\frac{\partial}{\p t}(A \circ B) = \frac{\partial A}{\p t}
\circ B + A \circ \frac{\partial B}{\p t}$. For stationary states,
wave functions simply change by a time-dependent phase and we can
study energy eigenvalues $E$ via the time-independent Schr\" odinger equation
\be \label{time-indep}
\mathcal H \star \psi = E \, \psi \ .
\ee

A Heisenberg-type approach focuses on $\star$-commutator based evolution equations for operators given by
\be
\frac{\partial A}{\p t} = \frac \ii\hbar \, [H,A]_\star \ .
\ee
This time evolution equation again fulfills all our requirements. It can be applied to single functions (``operators'') as well as to star products of functions, \emph{but it is not a derivation of~$\star$} since
\be
\begin{split}
[H,A]_\star \star B + A \star [H,B]_\star &= (H \star A)\star B - (A
\star H) \star B + A \star (H \star B) - A \star (B \star H)\\[4pt] 
&\neq [H,(A\star B)]_\star = H \star (A \star B) - (A\star B) \star H \ .
\end{split}
\ee
This surprising fact should be seen as an interesting feature of the theory, not as a mistake. We can still compute the time-dependence of any operator that we are interested in, but we cannot determine it from the time-dependence of its constituent parts. Similarly to the Schr\"odinger-type approach, there is an alternative equivalent way to compute the time-dependence of expectation values, in this case by the evolution equation for phase space state functions
\be
\frac{\partial S}{\p t} = \frac1{\ii\hbar} \, [H,S]_\star \ .
\ee
Stationary state functions $\star$-commute with the Hamiltonian function $H$ and we can study energy eigenvalues $E$ via
\be
H \star S = E \, S \ .
\ee
The only major difference between this equation and \eqref{time-indep} is that $S$ should be a real function while there is \emph{a priori} no such requirement for $\psi$.

We have found two sets of inequivalent but equally consistent transformation equations. Which approach should be used for what is ultimately a question of the physics that we would like to describe. The evolution equations of the Heisenberg-type approach close in the algebra of operators and appear therefore predestined to define active transformations like time evolution (i.e.\ dynamics), while the Schr\"odinger-type expressions could then still be useful to describe certain symmetries of the theory. In this section we have focused on single Hamiltonian dynamics instead of the also very interesting possibility of Nambu multi-Hamiltonian dynamics~\cite{Nambu:1973qe}.


\newsection{General $R$-fluxes\label{nonconstant}}

In this final section we briefly discuss some preliminary steps
towards extending the analysis of the present paper to more complicated
non-geometric $R$-flux compactifications. We consider, in particular,
the two separate cases in turn where the constant $3$-tensor $R^{ijk}$
is replaced with a general function of the position coordinates $x\in M$ and where
the $2$-tensor $\theta^{ij}(p) =R^{ijk}\, p_k$ is replaced by a
general function of the conjugate momenta
$p\in(\real^d)^*$. The former type of generalisation has been
discussed recently in
the context of double field theory in~\cite{Blumenhagen:2013hva},
while the latter type of generalisations arise
in~\cite{Lust:2010iy,Condeescu:2012sp}.


\subsection{Nambu-Poisson structures\label{NPstructures}}

The extension of our results to non-constant $R$-fluxes is closely related to the problem of
quantization of generic Nambu-Poisson structures (see
e.g.~\cite{DeBellis:2010pf, Saemann:2012ab} and references therein). A
Nambu-Poisson 3-bracket\footnote{There is a more general notion of
  Nambu-Poisson $p$-brackets for $p\geq2$ which however we do not need in this
  paper.} is a skew-symmetric ternary bracket defined on
the space of smooth functions $C^\infty(M)$ on a manifold $M$, which generalizes the
Poisson 2-bracket and can be expressed in terms of a trivector field
$\Pi \in C^\infty\big(M,\bigwedge^{3} TM \big)$ as $\{f,g,h\} =
\Pi(\dd f,\dd g,\dd h)$. The bracket is used to define a Nambu multi-Hamiltonian flow
\be
\frac{\dd f}{\dd t} = X_{H_1,H_2} f := \{f,H_1,H_2\}
\ee
with Nambu Hamiltonian vector field $X_{H_1,H_2}$ for any two smooth
functions $H_1$ and $H_2$. For a Nambu-Poisson structure, one requires that the vector fields $X_{H_1,H_2}$ act as a
derivation on the bracket, so that
\be \label{FundId}
X_{H_1,H_2} \{f,g,h\} = \{X_{H_1,H_2}f,g,h\} + \{f,X_{H_1,H_2}g,h\}+
\{f,g,X_{H_1,H_2} h\} \ .
\ee
This implies that the linear span of Nambu Hamiltonian vector fields defines a Lie algebra with Lie bracket
\be \label{Lieproperty}
[X_{f,g}, X_{f',g'}] = X_{X_{f,g}f'\,,\,g'}+
X_{f'\,,\,X_{f,g}g'} \ .
\ee
The condition \eqref{FundId}, when expressed solely in terms of brackets,
is known as the fundamental identity~\cite{Nambu:1973qe}. It is the
generalization of the Jacobi identity for Poisson brackets, which is a
differential condition on a Poisson bivector. For 3-brackets, the fundamental identity is a differential as well as an algebraic condition on the $3$-vector field $\Pi$. The algebraic condition implies~\cite{Guha96ondecomposability} that $\Pi$ is a decomposable trivector
\be
\Pi = X_1 \wedge X_2 \wedge X_3 \ .
\ee
The vector fields $X_1$, $X_2$ and $X_3$ are linearly independent (unless
$\Pi= 0$) and in view of \eqref{Lieproperty} they define an involutive
distribution. This implies that the local as well as the global
Frobenius theorem applies and in particular that around each point of
the manifold $M$ there exists a coordinate chart $(U;x^1,x^2,x^3)$ such that
\be \label{constantPi}
\Pi = \frac{\partial}{\partial x^1} \wedge \frac\partial{\partial x^2} \wedge \frac{\partial}{\partial x^3} \ .
\ee
This expression can be multiplied with a scalar pre-factor (e.g. a constant) without spoiling the properties of a Nambu-Poisson structure. 

The central object of interest of this paper is a constant trivector
$R$-flux $R = \frac1{3!} \, R^{ijk} \, \partial_i \wedge \partial_j \wedge \partial_k$. 
For appropriately chosen
coordinates, the decomposition \eqref{constantPi} implies that a
Nambu-Poisson tensor~$\Pi$ is in fact such a constant trivector, at
least locally, and most of the results of the present paper thus apply,
including the formalism of twist deformation quantization. Conversely, if $R$ extends locally in a three-dimensional submanifold of $M$, then $R$ is a Nambu-Poisson tensor. 
The parts of this paper dealing with integration apply to the
particular class of Nambu-Poisson structures for which \eqref{constantPi} holds globally.


\subsection{Non-parabolic monodromies \label{elliptic}}

When a three-torus $\torus^3$ in the $Q$-flux duality frame is viewed
as a $\torus^2$-fibration over $S^1$ as discussed in
Section~\ref{intro}, a periodic translation along the base must act on the
local complex structure modulus $\tau$ of a fibre $\torus^2$ as an
$SL(2,\zed)$ M\"obius transformation, in order to end up with an
automorphic fibre. These transformations define the monodromy
properties of the fibration and fall into conjugacy classes of
$SL(2,\zed)$~\cite{Hull:2005hk}; the case of trivial monodromies
corresponds to geometric spaces (manifolds), while non-trivial classes
correspond to non-geometric spaces (T-folds). Parabolic monodromies
are of infinite order and act as discrete shifts $\tau\mapsto \tau+n$,
where $n\in\zed$. As discussed in Section~\ref{intro}, under T-duality
the T-fold is mapped to the parabolic $R$-flux backgrounds
characterized by the phase space relations (\ref{luestalg}); this
algebra provides one of the simplest examples of nonassociativity and
may be regarded as the analog of the Moyal-Weyl background that arises
in open string theory with a constant $B$-field (see e.g.~\cite{Szabo:2001kg}). The case
of elliptic monodromies, which are of finite order and act as
$\zed_N$-transformations on the $\torus^2$ coordinates, were also
considered in~\cite{Lust:2010iy,Condeescu:2012sp} where it was shown
that the position coordinate commutator in (\ref{luestalg}) is
generalised to a particular non-linear function $\vartheta^{ij}(p)$ of momentum and
$R$-flux. We briefly describe here how to extend the setting of
Section~\ref{deformation} to allow for twist deformations that
correspond to a class of quasi-Poisson structures $\Theta_e$ which are generic functions of momentum. 

The class of generalizations of \eqref{Theta} that we are interested
in are obtained by substituting \eqref{theta} with a generic function
of momentum $\vartheta^{ij}(\tp)$ to get a bivector $\Theta_e= \frac12
\,\Theta_e^{IJ} \, \p_I \wg \p_J $ on phase space ${\cal M}$ given by 
\be
\Theta_e =\big(\Theta_e^{IJ} \big)= \begin{pmatrix} \vartheta^{ij} (p) & \delta^i{}_j \\
  -\delta_i{}^j & 0 \end{pmatrix} \ . \label{Thetae}
\ee
As in the case of the parabolic
$R$-flux model~\cite{Mylonas:2012pg}, a computation of the
Schouten-Nijenhuis bracket of this bivector with itself reveals that
it defines an $H$-twisted Poisson structure on $\Mcal$, where $H$ is
the $3$-form
\be
H=\mbox{$\frac16$} \, \tilde \p^i \vartheta^{jk}(p) \, \dd p_i \wg \dd p_j \wg \dd p_k
\ee
which is the curvature of a twisting $U(1)$ gerbe on momentum space.
The corresponding Jacobiator $J=\bigwedge^3 \Theta_e^\sharp (H)$ is
the $3$-vector whose only non-vanishing components are given by 
\be 
J^{ijk}(p) =\mbox{$\frac13$} \, \big(\tilde \p^i \vartheta^{jk}(p) + \tilde \p^j \vartheta^{ki}(p) + \tilde \p^k \vartheta^{ij}(p)\big) \ . \label{J}
\ee

Kontsevich's deformation quantization of a generic (quasi-)Poisson
structure is \emph{a priori} quite involved as the number of
weights that have to be calculated at each order of the
diagrammatic expansion of the star product increases geometrically. A
nonassociative star product up to third order in a derivative expansion of a generic
$B$-field was calculated in~\cite{Herbst:2003we} by using a twisted
Poisson sigma-model to determine the weights
of Kontsevich graphs; from the topological sigma-model formalism, the Kontsevich
formula inherits an invariance under the involution which interchanges
functions and maps $\Theta_e\mapsto-\Theta_e$. By applying the
open/closed string duality argument of~\cite{Mylonas:2012pg}, we can
transport their results to our closed string case and quantize the
$R$-flux background via the star product
\be
\begin{split}
f \star g = [f \,\ts\, g]_{\tp \to p}& - \mbox{$\frac{\hbar^2}{12}$} \,
\tilde \p^k \vartheta^{ij} \, \big( [ \p_k \p_i f\,\ts\, \p_jg]_{\tp \to
  p} + [\p_j f \,\ts\, \p_k \p_ig ]_{\tp \to p} \big) \\
& - \mbox{$\frac{\ii \hbar^3}{48}$} \, \tilde \p^l \tilde \p^k \vartheta^{ij} \, \big( [\p_l \p_k \p_if \,\ts\, \p_jg]_{\tp \to p} -[\p_j f\,\ts\, \p_l \p_k \p_ig]_{\tp \to p} \big) \\
& + \mbox{$\frac{\hbar^4}{288}$} \, \big(\tilde \p^l \vartheta^{mn}\big) \, \big(\tilde \p^k \vartheta^{ij}\big) \, \big( [\p_l \p_m \p_k \p_if \,\ts\, \p_n \p_jg]_{\tp \to p} \\ 
&  \qquad \qquad \qquad + 2 \, [\p_l \p_m \p_jf \,\ts\, \p_n \p_k \p_ig]_{\tp \to
  p} + [\p_n \p_jf \,\ts\, \p_l \p_m \p_k \p_ig]_{\tp \to p} \big) \\ & +
{\cal O}\big(\tilde\p^3 \vartheta, (\tilde \p \vartheta)^3 \big) \label{esp}
\end{split}
\ee
for $f,g \in C^\infty ({\cal M})$, where as before the operation $[-]_{\tp \to p}$ denotes the change from constant to dynamical momentum and
\be
f \,\ts\, g = \mu_A \Big( \exp\big[\mbox{$\frac{\ii \hbar}{2}$} \big(
\vartheta^{ij}(\tp)\, \p_i \ot \p_j + \p_i \ot \tilde \p^i - \tilde
\p^i \ot \p_i \big) \big] (f\ot g) \Big) 
\ee
is an associative Moyal-Weyl type product on $C^\infty ({\cal
  M})$. One can check that this star product is
nonassociative and that it reduces to the star
product \eqref{sp} in the parabolic case $\vartheta^{ij}(p) =
\theta^{ij}(p)=R^{ijk}\, p_k$, by antisymmetry of the $R$-flux
components. In particular, by substituting $f$ and $g$ with phase
space coordinates we find the quantum phase space relations
\be 
[x^i,x^j]_\star=\ii \hbar\, \vartheta^{ij} (p) \ ,\qquad
[x^i,p_j]_\star = \ii \hbar \, \delta^i{}_j \ , \qquad [p_i,p_j]_\star=0 \label{qealg}
\ee
while the quantized Jacobiator is
\be
[[ x^i, x^j, x^k]]_\star = 3\, \hbar^2 \, J^{ijk}(p) \ . \label{qejac}
\ee 

We may now construct the pertinent Hopf algebra of symmetries of the
closed string background. Consider the non-abelian
Lie algebra $\tmfg$ generated by $\tP^i$ and $P_i^{(f)}
:=f(p)\,P_i$, where $f(p) \in C^\infty( (\real^d)^* )$, with the only
non-trivial commutation relations given by
\be 
\big[\tP^i \,,\, P_j^{(f)} \big] =  P^{(\tilde \p^i f )}_j \ . \label{ecom}
\ee
The generators of $\tmfg$ are realised on phase space ${\cal M}$ by the action
\eqref{rep} on
$C^\infty(\Mcal)$; they respectively generate momentum translations and
position translations together with a momentum-dependent scaling by $f(p)$. In
particular, this infinite-dimensional Lie algebra contains the Lie
subalgebra of translations and Bopp shifts in phase space $\mfg$ that we used in Section~\ref{deformation}. 
The pertinent Hopf algebra $H_e$ is the universal enveloping algebra $U(\tmfg)$ equiped with the coalgebra structure \eqref{costructure}.

A suitable (but not unique) twist $\mcF_e \in H_e[[\hbar]] \ot
H_e[[\hbar]]$ that
reproduces the star product (\ref{esp}) is given by 
\be
\begin{split}
\mcF_e = [\tmcF_e]_{\tp \to p} & - \mbox{$\frac{\hbar^2}{12}$} \,
\tilde \p^k \vartheta^{ij}(p) \, [\tmcF_e]_{\tp \to p} \, \big( P_k \, P_i \,\ot\, P_j + P_j \,\ot\, P_k \, P_i \big) \\
& - \mbox{$\frac{\ii \hbar^3}{48}$} \, \tilde \p^l \tilde \p^k \vartheta^{ij}(p) \, [\tmcF_e]_{\tp \to p} \, \big( P_l\, P_k\, P_i \,\ot\, P_j - P_j \,\ot\, P_l\, P_k\, P_i \big) \\
& + \mbox{$\frac{\hbar^4}{288}$} \, \big(\tilde \p^l \vartheta^{mn}(p)\big) \, \big(\tilde \p^k \vartheta^{ij}(p)\big) \, [\tmcF_e]_{\tp \to p} \ \big( P_l\, P_m\, P_k\, P_i \,\ot\, P_n\, P_j \\
& \qquad \qquad \qquad \qquad +2 \, P_l\, P_m\, P_j \,\ot\, P_n\, P_k\, P_i + P_n\, P_j
\,\ot\, P_l\, P_m\, P_k\, P_i \big) \\ & +
{\cal O}\big(\tilde\p^3 \vartheta, (\tilde \p \vartheta)^3 \big) \ , \label{emcFe}
\end{split}
\ee
where 
\be
\tmcF_e = \exp{\big[ \mbox{$-\frac{\ii\hbar}{2}$}\, \big( \vartheta^{ij}(\tp) \, P_i \,\ot\, P_j + P_i \,\ot\, \tilde P^i - \tilde P^i \,\ot\, P_i\big)\big]} \ . \label{tmcFe}
\ee
The expression \eqref{emcFe} defines a 2-cochain on $H_e[[\hbar]]$ with
coboundary $\phi_e=\partial^*\mcF_e$ given by
\be
\begin{split}
\phi_e =& \, 1\,\ot\, 1 \,\ot\, 1 + \mbox{$\frac{\hbar^2}{2}$} \, J^{ijk}(p) \, [\tmcF_{12} \, \tmcF_{23}]_{\tp \to p} \, (P_i \,\ot\, P_j \,\ot\, P_k ) \\ 
& + \mbox{$\frac{\ii \hbar^3}{8}$} \, \tilde \p^l J^{ijk}(p) \, [\tmcF_{12} \, \tmcF_{23}]_{\tp \to p} \, \big( P_l\, P_i \,\ot\, P_j \,\ot\, P_k - P_j \,\ot\, P_k \,\ot\, P_l\, P_i \big) \\
& + \mbox{$\frac{ \hbar^4}{8}$} \, \big( \tilde \p^l \vartheta^{mn}(p) \big) \, J^{ijk}(p) \, [\tmcF_{12} \, \tmcF_{23}]_{\tp \to p} \\ 
& \qquad \times \big( P_l\, P_m\, P_i \,\ot\, P_k \,\ot\, P_n\, P_j + P_l\, P_m\, P_i \,\ot\, P_n\, P_k \,\ot\, P_j + P_i\, P_n \,\ot\, P_l\, P_m\, P_k \,\ot\, P_j \\  
& \qquad\qquad + P_i \,\ot\, P_l\, P_m\, P_k \,\ot\, P_n\, P_j + 2 \, P_l\, P_i \,\ot\, P_m\, P_k \,\ot\, P_n\, P_j + 2\, P_n\, P_i \,\ot\, P_m\, P_k \,\ot\, P_l\, P_j \\ 
& \qquad\qquad + P_i \,\ot\, P_n\, P_k \,\ot\, P_l\, P_m\, P_j + P_i\, P_n \,\ot\,
P_k \,\ot\, P_l\, P_m\, P_j \big) \\ & + {\cal O}\big(\tilde\p^3
\vartheta, (\tilde \p \vartheta)^3 \big) \ . \label{phi_e}
\end{split}
\ee
It is straightforward to check that the expression \eqref{phi_e}
satisfies the cocycle condition \eqref{3cocycle} order by order in $\hbar$, and hence yields
the counital associator 3-cocycle for the quasi-Hopf algebra obtained from
twisting $H_e$ by $\mcF_e$. Note that each term in (\ref{phi_e})
involves the classical Jacobiator \eqref{J}. 

The generality of this setting now allows
for deformation quantization of the geometry of elliptic $R$-flux
backgrounds up to third order in the $R$-flux. However, in this case
cyclicity of the nonassociative star product \eqref{esp} is a more
delicate issue and requires a more sophisticated definition of
integration on non-parabolic $R$-spaces; see~\cite{Herbst:2003we} for
a detailed analysis of this problem in the context of open string
theory, and~\cite{Blumenhagen:2013hva} for an investigation in the
context of double field theory.

\subsection*{Acknowledgments}

We thank P.~Aschieri, R.~Blumenhagen, A.~Chatzistavrakidis, C.-S.~Chu,
J.~Figueroa-O'Farrill, O.~Lechtenfeld, F.~Lizzi, D.~L\"ust, R.~Nest and C.~S\"amann for helpful
discussions. A preliminary version of this paper was presented by
D.M. at the Workshop on ``Noncommutative Field Theory and Gravity''
which took part from 8--15 September 2013 as part of
the Corfu Summer Institute on Elementary Particle Physics and Gravity. The work of
D.M. is supported by 
the Greek National Scholarship Foundation. The work of D.M. and R.J.S. was supported in part by the Consolidated Grant ST/J000310/1
from the UK Science and Technology Facilities Council. P.S. thanks the Deutsche Forschungsgemeinschaft for support within the RTG
1620 ``Models of Gravity'', and P.-M.~Ho and the CTS at NTU, Taipei, for
helpful discussions, hospitality, and support.

\end{document}